\documentclass[11pt]{article}

\usepackage{epsfig}
\usepackage{latexsym}
\usepackage{graphicx}
\usepackage{color}
\usepackage{amsmath,amssymb}
\usepackage{cite}

\makeatletter
\def\section{\@startsection {section}{1}{\z@}{-3.5ex plus -1ex minus -.2ex}{2.3ex plus .2ex}{\large\bf}}
\def\subsection{\@startsection{subsection}{2}{\z@}{-3.25ex plus -1ex
minus -.2ex}{1.5ex plus .2ex}{\normalsize\bf}}
\makeatother
\newcommand{\captionfonts}{\small}
\makeatletter  
\long\def\@makecaption#1#2{%
  \vskip\abovecaptionskip
  \sbox\@tempboxa{{\captionfonts #1: #2}}%
  \ifdim \wd\@tempboxa >\hsize
    {\captionfonts #1: #2\par}
  \else
    \hbox to\hsize{\hfil\box\@tempboxa\hfil}%
  \fi
  \vskip\belowcaptionskip}
\makeatother   

\catcode`@=11
\def\marginnote#1{}
\newcount\hour
\newcount\minute
\newtoks\amorpm
\hour=\time\divide\hour
by60
\minute=\time{\multiply\hour by60 \global\advance\minute
by-\hour}
\edef\standardtime{{\ifnum\hour<12 \global\amorpm={am}
\else\global\amorpm={pm}\advance\hour by-12 \fi
 \ifnum\hour=0
\hour=12 \fi
 \number\hour:\ifnum\minute<10
0\fi\number\minute\the\amorpm}}
\edef\militarytime{\number\hour:\ifnum\minute<10
0\fi\number\minute}
\def\draftlabel#1{{\@bsphack\if@filesw
{\let\thepage\relax
 \xdef\@gtempa{\write\@auxout{\string
\newlabel{#1}{{\@currentlabel}{\thepage}}}}}\@gtempa
 \if@nobreak
\ifvmode\nobreak\fi\fi\fi\@esphack}
\gdef\@eqnlabel{#1}}
\def\@eqnlabel{}
\def\@vacuum{}
\def\draftmarginnote#1{\marginpar{\raggedright\scriptsize\tt#1}}
\def\draft{\oddsidemargin
0.0truein
 \def\@oddfoot{\sl preliminary draft \hfil
\rm\thepage\hfil\sl\today\quad\militarytime}
 \let\@evenfoot\@oddfoot
\overfullrule 3pt
 \let\label=\draftlabel
\let\marginnote=\draftmarginnote
\def\@eqnnum{(\theequation)\rlap{\kern\marginparsep\tt\@eqnlabel}
\global\let\@eqnlabel\@vacuum}
}
\catcode`@=12

\newcommand{\beq}{\begin{eqnarray}}
\newcommand{\eeq}{\end{eqnarray}}
\newcommand{\f}[2]{\frac{#1}{#2}}

\newcommand{\gsim}{\raisebox{-0.13cm}{~\shortstack{$>$ \\[-0.07cm]
     $\sim$}}~}
\newcommand{\lsim}{\raisebox{-0.13cm}{~\shortstack{$<$ \\[-0.07cm]
     $\sim$}}~}
\newcommand{\s}{\newline \vspace*{-3.5mm}}

\begin{document}

\thispagestyle{empty}

\begin{center}
\hfill KA-TP-22-2014 \\
\hfill PSI-PR-14-09 \\
\hfill SFB/CPP-14-62 \\

\begin{center}

\vspace{1.7cm}

{\LARGE\bf Light Stop Decays: \\[0.2cm] Implications for LHC
  Searches}
\end{center}

\vspace{1.4cm}

\renewcommand{\thefootnote}{\fnsymbol{footnote}}
{\bf R.~Gr\"ober$^{\,1\,}$}\footnote{E-mail: \texttt{ramona.groeber@kit.edu}}, {\bf M. M\"uhlleitner$^{\,1}\,$}\footnote{E-mail: \texttt{margarete.muehlleitner@kit.edu}}, {\bf E. Popenda$^{\,2\,}$}\footnote{E-mail: \texttt{eva.popenda@psi.ch}} and {\bf A.~Wlotzka$^{\,1\,}$}\footnote{E-mail: \texttt{alexander.wlotzka@kit.edu}}\\

\vspace{1.2cm}

${}^1\!\!$
{\em Institute for Theoretical Physics, Karlsruhe Institute of
    Technology,} \\
{\em 76128 Karlsruhe, Germany}
\\
${}^2\!\!$
{\em {Paul Scherrer Institut, CH-5232 Villigen PSI, Switzerland}
}\\

\end{center}

\vspace{1.8cm}
\centerline{\bf Abstract}
\vspace{2 mm}
\begin{quote}
\small We investigate the flavour-changing neutral current decay of
the lightest stop into a charm quark and the lightest neutralino
and its four-body decay into the lightest neutralino, a down-type quark and
a fermion pair. These are the relevant stop search channels in the
low-mass region. The SUSY-QCD corrections to the two-body decay have
been calculated for the first time and turn out to be sizeable. In the
four-body decay both the contributions from diagrams with
flavour-changing neutral current (FCNC) couplings and the mass effects
of  final state bottom quarks and $\tau$ leptons have been taken into
account, which are not available in the literature so far. The resulting
branching ratios are investigated in detail. We find that in
either of the decay channels the branching ratios can deviate
significantly from one in large parts of the allowed parameter
range. Taking this into account, the experimental exclusion limits on the stop,
which are based on the assumption of branching ratios equal to one, are
considerably weakened. This should be taken into account in future
searches for light stops at the next run of the LHC, where the probed
low stop mass region will be extended. 
\end{quote}

\newpage
\setcounter{page}{1}
\setcounter{footnote}{0}

\section{\label{sec:intro} Introduction}
With the discovery of a new scalar particle by the LHC experiments
ATLAS and CMS \cite{:2012gk,:2012gu} we have entered a new era of particle
physics. The investigation of its properties, like spin and CP
quantum numbers and couplings to other Standard Model (SM) particles, have
identified it as the long-sought Higgs particle predicted by the
Higgs mechanism \cite{higgsmech}. The absence of any discovery of new
particles beyond the SM, however, leaves the question of the
underlying dynamics of the mechanism of electroweak symmetry breaking
open. Models with the Higgs boson emerging as composite bound 
state from a strongly coupled sector \cite{strong} are compatible with
the LHC data, as well as extensions like supersymmetry (SUSY)
\cite{susy} based on a weakly interacting theory. One of the main
goals of the LHC is therefore the search for new particles and the
subsequent investigation of their properties in order to pin down the true nature
of the discovered Higgs boson. \s 

Among the plethora of beyond the SM (BSM) extensions, SUSY is
one of the most extensively studied models. It requires the
introduction of at least two complex Higgs doublets, leading in its
most economic version, the Minimal Supersymmetric
Extension of the SM (MSSM) \cite{mssm}, to five Higgs bosons, among which
the lightest CP-even state $h$ can be identified with the recently
discovered SM-like boson. Within SUSY models the
hierarchy problem can be solved by the symmetry between bosonic and
fermionic degrees of freedom. Assuming SUSY to be softly broken, the
Higgs mass corrections grow logarithmically with the square of the
SUSY scale $m_S$. The loop corrections from the top loops and their
SUSY partners, the stops, are crucial in order to shift the mass of
the lightest SUSY Higgs boson above the upper tree-level bound set by
the $Z$ boson mass $M_Z$. With the SUSY scale given by the
average stop mass, $m_S^2=m_{\tilde{t}_1} m_{\tilde{t}_2}$, and the
stop mixing parameter $X_t$, the mass squared of the lightest Higgs
boson including the leading corrections in the SM limit, is given by
\beq
M_h^2 = M_Z^2 \cos^2 2\beta + \frac{3 m_t^4}{2 \pi^2 v^2} \left( \log
  \left( \frac{m_S^2}{m_t^2} \right) + X_t^2 \left( 1 -
    \frac{X_t^2}{12} \right) \right)
\;,
\eeq
with $m_t$ denoting the top quark mass, $v$ the vacuum expectation
value (VEV) with $v \approx 246$~GeV and
\beq
X_t = \frac{A_t - \mu \cot \beta}{m_S} \;.
\eeq 
The ratio of the two VEVs of the
neutral components of the MSSM Higgs doublets is given by
$\tan\beta$ and $A_t$ denotes the soft SUSY breaking trilinear
coupling in the stop sector. 
A large Higgs boson mass of around 125~GeV can hence be
obtained either through a large stop mixing $X_t$ or through heavy
stops. Naturalness arguments suggest the stops to be light, since the
amount of fine-tuning of the electroweak scale is significantly driven by the
stop \cite{Dimopoulos:1995mi}. The maximal mixing scenario, with
$X_t^2 \approx 6$ and $m_S \approx 500$~GeV leading to the observed
Higgs mass value, therefore optimally reduces the amount of
fine-tuning \cite{Wymant:2012zp}. In most SUSY models a light stop
arises naturally due to the mixing being proportional to the large
Yukawa coupling, which leads to a large mass splitting between the stop
mass eigenstates. \s

Light stops not only play a special role in view of the Higgs mass and
naturalness arguments. A light stop can also lead to the correct relic
density through co-annihilation, in particular for mass differences
between the stop and the lightest neutralino $\tilde{\chi}_1^0$
of $15-30$~GeV or a pseudoscalar mass $M_A$ with $M_A\approx 2
m_{\tilde{\chi}_1^0}$ \cite{Boehm:1999bj}. Moreover, light stops allow for
successful baryogenesis within the MSSM \cite{EWBG}.\footnote{This
  requires, however, a stop mass of about the top mass value or below,
which is in tension with the experimental direct stop search limits,
see {\it e.g.}~\cite{atlas1,atlas2,atlas3} and limits from the
measurement of the $t\bar{t}$ cross section
\cite{Aad:2014kva,Czakon:2014fka}.} \s 

Despite the LHC searches pushing the limits on the copiously produced
coloured sparticles above the $1-1.5$ TeV range
for the first two generations \cite{atlassusybound,cmssusybound}, the
lightest stop can still be rather light, 
with masses below the kinematical thresholds for the decay into a top
and a lightest neutralino $\tilde{\chi}_1^0$, $\tilde{t}_1 \to t
\tilde{\chi}_1^0$, and for the decay $\tilde{t}_1 \to \tilde{\chi}_1^0
W b$ into a neutralino, a $W$ boson and a bottom quark $b$. Assuming
the lightest stop to be the next-to-lightest  
supersymmetric particle and the $\tilde{\chi}_1^0$ to be the
lightest SUSY particle (LSP), the light stop can then decay into the
LSP and a charm quark $c$ or an up quark $u$, $\tilde{t}_1 \to (u/c)
\tilde{\chi}_1^0$ \cite{Hikasa:1987db,Muhlleitner:2011ww}. Another
possible decay channel is the four-body decay $\tilde{t}_1 \to 
\tilde{\chi}_1^0 b f \bar{f}'$ \cite{Boehm:1999tr}, with $f$ and $f'$
denoting generic light fermions. The two-body decay into charm/up and
neutralino is flavour-violating (FV). The MSSM in general exhibits many
sources of flavour violation, so that the decay can already occur at
tree-level. High precision tests in the sector of quark flavour
violation and limits on flavour-changing neutral currents from
$K$, $D$ and $B$ meson studies put stringent constraints on the amount of
possible flavour violation \cite{Grossman:2009dw}. In order to
solve this New Physics Flavour Puzzle the framework of Minimal Flavour
Violation (MFV) has been proposed
\cite{Chivukula:1987fw,Buras:2000dm,D'Ambrosio:2002ex,Bobeth:2005ck}, which
requires all sources of flavour and CP violation to be given by the SM
structure of the Yukawa couplings. The hypothesis of MFV is not
renormalisation group invariant \cite{D'Ambrosio:2002ex}, however,
inducing flavour off-diagonal squark mass terms through the Yukawa
couplings, which results in tree-level FCNC couplings. If the FV 
stop-neutralino-up/charm quark coupling is very small, the four-body
decay can become important and has to be taken into account for a
reliable prediction of the $\tilde{t}_1$ branching ratios. \s

Bounds on the stop masses have been set by LEP \cite{lepsearch}  and
Tevatron \cite{tevsearch}, and more recently by the ATLAS \cite{atlas1} and
the CMS \cite{cms1} collaborations. The strongest
limits come from the ATLAS analyses Refs.~\cite{atlas2,atlas3}. 
All these analyses assume a branching
ratio of one for the analysed decay channel of the $\tilde{t}_1$, either the
FV two-body or the four-body decay.
However, in Ref.~\cite{Muhlleitner:2011ww} it was already pointed out
that the competing FV two-body and four-body stop
decays can lead to substantial deviations from branching ratios of
one in either of the decay channels. This has a significant impact on the
stop mass bounds set by the experiments. The calculation in
Ref.~\cite{Muhlleitner:2011ww} improved the existing approximate
result for the $\tilde{t}_1 \to (u/c) \tilde{\chi}_1^0$ decay of
Ref.~\cite{Hikasa:1987db} by computing the exact one-loop decay width in the
framework of MFV. Resummation effects, that can 
become important, have not been taken into account in that
approach. In this work, we therefore include resummation
effects through renormalisation group running induced FCNC couplings
already at tree-level and calculate the one-loop SUSY-QCD corrections
to the two-body decay. In order to correctly 
determine the $\tilde{t}_1$ branching ratios, also the four-body decay is
computed by consistently including FCNC couplings. Moreover,
non-vanishing masses for the third generation final state fermions
have been taken into account. These decay widths have been
implemented in the Fortran code {\tt SUSY-HIT} \cite{susyhit} for the
calculation of the decay widths and branching ratios of SUSY particles
in the MSSM. With the thus obtained $\tilde{t}_1$
branching ratios we discuss the implications for the LHC stop searches
and the bounds obtained on the mass of the lightest
stop $m_{\tilde{t}_1}$. The program with the newly implemented stop decays is
available at \cite{program}. \s

The outline of the paper is as follows. In Section \ref{sec:loopcorr}
we present the calculation of the SUSY-QCD corrections to the FCNC
two-body decay. The computation of the four-body decay is deferred to
Section \ref{sec:4body}. In Section~\ref{sec:parameterscan} the details
of our parameter scan are given as well as the applied constraints. We
discuss our results in the numerical analysis in
Section~\ref{sec:numerical}. Section~\ref{sec:concl} summarises our findings. 

\section{\label{sec:loopcorr} The Flavour-Violating
  Two-Body Stop Decay}
The two-body decay of the lightest stop into a charm or an up quark
and the lightest neutralino is mediated at tree-level by a FCNC
coupling. In the MSSM with flavour violation the squark and
quark mass matrices cannot be diagonalised simultaneously any more.
The squarks are no longer flavour eigenstates and the SUSY partners of
the left- and right-chiral up-type quarks mix to form a
six-component vector $\tilde{u}_s$ ($s=1,...,6$). Analogously, the
down-type squarks are described by the six-component vector
$\tilde{d}_s$. We assume the entries to be ordered in mass, with
$\tilde{u}_1$ ($\tilde{d}_1$) denoting the lightest up-type
(down-type) squark. The MFV approach naturally accounts for small flavour
violation, with the only source of flavour violation being the CKM
matrix. A way to implement it, is by assuming that the squark and
quark mass matrices can be diagonalised simultaneously at a scale $\mu =
\mu_{\text{MFV}}$, so that there are no FCNC couplings at tree
level. Flavour mixing is induced through renormalisation group
equation (RGE) running at any scale $\mu \ne \mu_{\text{MFV}}$. Due to
the large mixing in the stop sector, the lightest up-type squark $\tilde{u}_1$ is
hence mostly stop-like. In the following, we will refer to
$\tilde{u}_1$ as the lightest stop where appropriate, although it is
understood that it has a small flavour admixture from the charm- and
up-flavours. Considering a light stop with a mass close to the one of
the lightest neutralino, it mainly decays through the FV two-body decays
\beq 
\tilde{u}_1 \to (u/c) + \tilde{\chi}_1^0 \;. \label{eq:fl2bod}
\eeq
Due to the smallness of the up-flavour admixture (because of the
small CKM matrix elements, which are responsible for flavour mixing
through RGE running) the decay into the up quark final state is
suppressed by about two orders of magnitude compared to the charm quark final
state. We have performed our calculations for both final states, but
will discuss here the one with the charm quark in the final state.

\subsection{The Squark Sector}
In order to set up our notation we start with the introduction of the
squark sector. Denoting by $\tilde{q}_L^\prime$ and
$\tilde{q}_R^\prime$, respectively, a three-component vector in
generation space, we define the six-component vector
$\tilde{q}^\prime$ describing the squark interaction eigenstates,
\beq
\tilde{q}^\prime = \left( \begin{array}{c} \tilde{q}_L^\prime \\ \tilde{q}_R^\prime
\end{array} \right) \;.
\eeq
The squark mass matrix, written as a $2\times 2$ Hermitian matrix of
$3\times 3$ blocks,
\beq
{\cal M}_{\tilde{q}^\prime}^2 = \left( \begin{array}{cc} {\cal
    M}_{\tilde{q}^\prime_{LL}}^2 &   {\cal M}_{\tilde{q}^\prime_{LR}}^2 \\
 {\cal M}_{\tilde{q}^\prime_{RL}}^2 &  {\cal
    M}_{\tilde{q}^\prime_{RR}}^2 \end{array} \right) \; ,
\eeq
is diagonalised by a $6 \times 6$ unitary matrix $\widetilde{W}$,
rotating the squark interaction eigenstates to the mass eigenstates
$\tilde{q}^m$,
\beq
\tilde{q}^m = \widetilde{W} \tilde{q}^\prime \;,
\eeq
where the $\tilde{q}^m$ are ordered in mass. We can decompose the
squark mass eigenstate field into left- and right-chiral interaction
eigenstates through ($s=1,...,6$, $i=1,2,3$)
\beq
\tilde{q}^m_s = \widetilde{W}_{si} \tilde{q}^\prime_{iL}  + \widetilde{W}_{s
  \, i+3} \tilde{q}^\prime_{iR} \equiv (\widetilde{W}_L \tilde{q}^\prime_L
+ \widetilde{W}_R \tilde{q}^\prime_R)_s \;, \label{eq:sq1}
\eeq
where $i$ is the generation index. The matrices $U^{u_{L,R}}$ and
$U^{d_{L,R}}$ are the $3\times 3$ unitary matrices that rotate the left- and
right-handed up- and down-type current eigenstates $u_{L,R}$ and
$d_{L,R}$ to their corresponding mass eigenstates, $u^m_{L,R}$ and
$d^m_{L,R}$,
\beq
u_{L,R}^m = U^{u_{L,R}} u_{L,R} \qquad \mbox{and} \qquad
d_{L,R}^m = U^{d_{L,R}} d_{L,R} \; . \label{eq:sq2}
\eeq 
They define the CKM matrix $V^{\text{CKM}}$ as
\beq
V^{\text{CKM}} = U^{u_L} U^{d_L\dagger} \;. 
\eeq
In the super-CKM basis the squarks are rotated by the same unitary
matrices as the quarks, implying that at scales $\mu \ne
\mu_{\text{MFV}}$ or in non-minimal flavour violation models, the
squark mass matrix is flavour-mixed, contrary to the quark mass
matrix. Otherwise, the squarks are flavour eigenstates after rotation
by $U^{q_{L,R}}$, and we have 
\beq
\tilde{q}_L = U^{q_L} \tilde{q}_L^\prime \qquad \mbox{and} \qquad
\tilde{q}_R = U^{q_R} \tilde{q}_R^\prime  \;, \label{eq:qrotation}
\eeq
with the squared mass matrix in the flavour eigenstate basis
$(\tilde{q}_L,\tilde{q}_R)^T$ given by
\beq
{\cal M}_{\tilde{q}}^2 = \left( \begin{array}{cc}
    (\tilde{M}_{\tilde{q}_L}^2 +m_q^2) {\bf 1}_3 & m_q (A_q -\mu r_q)
    {\bf 1}_3 \\ m_q (A_q -\mu r_q) {\bf 1}_3 &
    (\tilde{M}_{\tilde{q}_R}^2 +m_q^2) {\bf 1}_3 
 \end{array} \right) \;,
\eeq
where ${\bf 1}_3$ denotes a $3\times 3$ unit matrix. 
Here $\tilde{M}_{\tilde{q}_{L,R}}$ are given by the left- and
right-handed scalar soft SUSY breaking masses $M_{\tilde{q}_{L,R}}$
and the $D$-terms
\beq
\tilde{M}_{\tilde{q}_{L,R}}^2 &=& M_{\tilde{q}_{L,R}}^2 +
D_{\tilde{q}_{L,R}} \\
D_{\tilde{q}_L} &=& M_Z^2 \cos 2\beta (I^3_q - Q_q \sin^2 \theta_W) \\
D_{\tilde{q}_R} &=& M_Z^2 \cos 2\beta Q_q \sin^2 \theta_W \;,
\eeq
with the third component $I^3_q$ of the weak isospin of the quark $q$,
$Q_q$ its electric charge and $\theta_W$ denoting the Weinberg
angle. The soft SUSY breaking trilinear coupling is given by $A_q$, 
and $\mu$ stands for the higgsino mass parameter. In addition, we have
used the abbreviations $r_d= 1/r_u = \tan\beta$ for down- and
up-type quarks. The flavour eigenstates are rotated to their mass
eigenstates by the $6\times 6$ unitary matrix $W$, $(s,t=1,...,6$,
$i=1,2,3)$, 
\beq
\tilde{q}^m_s = W_{st} \left( \begin{array}{c} \tilde{q}_L \\
    \tilde{q}_R \end{array} \right)_t = W_{si} \, \tilde{q}_{Li} +
W_{s \, i+3} \, \tilde{q}_{Ri} \equiv (W_L \tilde{q}_L + W_R
\tilde{q}_R)_s \;.
\label{eq:sq3}
\eeq
The $6\times 3$ matrices $\widetilde{W}_{L,R}$ can hence be factorised
into the $6\times 3$ matrices $W_{L,R}$, which are flavour-diagonal at
$\mu_{\text{MFV}}$, and the $3\times 3$ quark rotation matrices,
\beq
\widetilde{W}_L = W_L U^{q_L}  \qquad \mbox{and} \qquad \widetilde{W}_R = W_R
U^{q_R} \;, \qquad q=u,d\;, 
\eeq
as can be inferred from comparing Eq.~(\ref{eq:sq3}) with
Eq.~(\ref{eq:sq1}) and using Eq.~(\ref{eq:qrotation}). \s

\subsection{The Loop-Corrected Stop Two-Body Decay}
Defining the $4\times 4$ neutralino mixing matrix $Z$ diagonalising the
neutralino mass matrix in the bino, wino, down- and up-type higgsino
basis $(-i\tilde{B},-i\widetilde{W}_3, \tilde{H}_1^0,
\tilde{H}_2^0)$, we can write the coupling between an up-type quark
$u_i$ ($i=1,2,3$), an up-type squark $\tilde{u}_s$ ($s=1,...,6$) and a neutralino
$\tilde{\chi}^0_l$ ($l=1,...,4$) in terms  of the left- and right-chiral 
couplings as
\beq
g^L_{isl} &=& -ge^{u_i}_{Rl} W^\dagger_{i+3 \,s} - \frac{gZ_{l4} m_{u_i}
\delta_{ij}}{\sqrt{2} M_W \sin\beta} W^\dagger_{js} \label{eq:leftchiral} \\
g^R_{isl} &=& -ge^{u_i}_{Ll} W^\dagger_{is} - \frac{g Z_{l4} m_{u_i}
  \delta_{ij}}{\sqrt{2} M_W \sin\beta} W^\dagger_{j+3 \, s} \;. \label{eq:rightchiral}
\eeq
Here $M_W$ and $m_{u_i}$ denote, respectively, the mass of the $W$
boson and of the quark and $g$ the $SU(2)$ gauge
coupling. Furthermore, we have introduced the abbreviations
\beq
e^q_{Ll} &=& \sqrt{2} [Z_{l1} t_W (Q_q-I^3_q)+Z_{l2} I_q^3]  \\
e^q_{Rl} &=& -\sqrt{2} Q_q t_W Z_{l1} \;,
\eeq 
where $t_W$ is a short-hand notation for $\tan\theta_W$. We
can then write the leading order tree-level two-body decay width for the decay of
the lightest up-type squark into a charm quark and the lightest
neutralino as
\beq
\Gamma^{\text{LO}} (\tilde{u}_1 \to c \tilde{\chi}_1^0) =
\frac{m_{\tilde{u}_1}}{16\pi}
\lambda(m_{c}^2,m_{\tilde{\chi}_1^0}^2;m_{\tilde{u}_1}^2) &&\hspace*{-0.7cm}\left[ -4
  g^L_{211} g^R_{211} \frac{m_c
    m_{\tilde{\chi}_1^0}}{m_{\tilde{u}_1}^2} \right. \nonumber \\
&& \hspace*{-0.7cm}+ \left. \left( 1-
   \frac{m_c^2+m_{\tilde{\chi}_1^0}^2}{m_{\tilde{u}_1}^2}\right)\left(
      (g^L_{211})^2 + (g^R_{211})^2 \right) \right]  \; , \label{eq:lowidth}
\eeq  
with the two-body phase space function
\beq
\lambda(x,y;z) = \sqrt{(1-x/z-y/z)^2 - 4xy/z^2} 
\eeq
and the lightest up-type squark and neutralino masses,
$m_{\tilde{u}_1}$ and $m_{\tilde{\chi}_1^0}$. Note, that for a non-vanishing
decay width the flavour off-diagonal matrix elements of the squark mixing
matrix $W$ have to be non-vanishing. For simplicity, in the following
we set the charm quark mass to zero, which does not have any
significant effects unless the mass difference between the decaying squark
and the neutralino becomes comparable with the charm quark mass or 
the lightest neutralino becomes mostly higgsino-like. For a mass
difference of 5~GeV {\it e.g.}~the difference between the leading
order (LO) decay width with $m_c=0$ and the one with non-vanishing
charm quark mass is about 3\%  and less than 1\% for 10~GeV mass
difference. In mSUGRA models the lightest neutralino 
for a top quark mass of 173~GeV never becomes higgsino-like
\cite{dreesnojiri}. The lightest neutralino can only be higgsino-like
for mass values close to the mass of the lightest chargino. While the
limits on the chargino masses are model-dependent, the parameter space for
light charginos gets more and more constrained  by the LHC experiments
\cite{charginolimits1,charginolimits2}. In scenarios with the lightest
neutralino mass much lighter than the chargino masses, the neutralino is mainly
gaugino-like. \s

The decay width $\Gamma^{\text{NLO}}$ including the next-to-leading
order (NLO) SUSY-QCD corrections is composed of the LO
width $\Gamma^{\text{LO}}$, of the contributions $\Gamma^{\text{virt}}$
from the virtual corrections, $\Gamma^{\text{real}}$ from the real corrections and
the one arising from the counterterms, $\Gamma^{\text{CT}}$, 
\beq
\Gamma^{\text{NLO}} = \Gamma^{\text{LO}} + \Gamma^{\text{virt}} +
\Gamma^{\text{real}} + \Gamma^{\text{CT}} \;. \label{eq:nlowidth}
\eeq

\subsubsection{The NLO SUSY-QCD Corrections} 
\begin{figure}[t]
\begin{center}
\includegraphics[width=8cm]{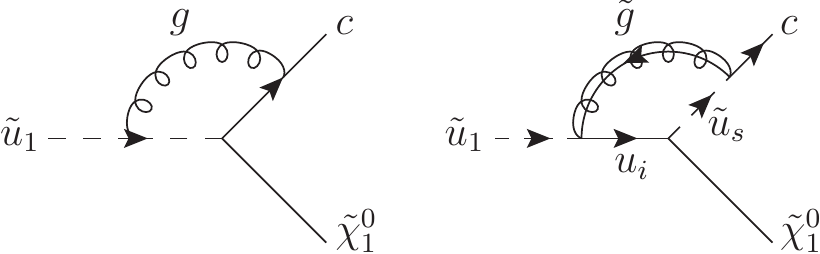} \\[0.7cm]
\includegraphics[width=12.5cm]{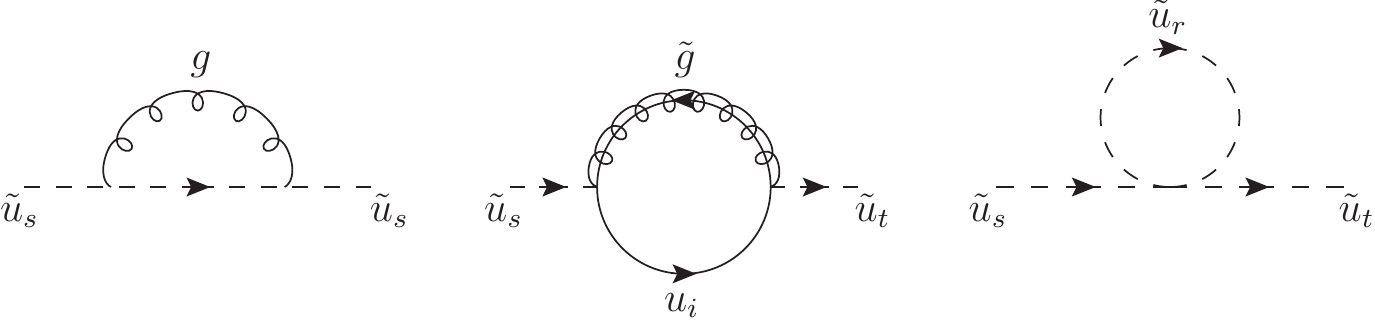} \\[0.7cm]
\includegraphics[width=8.4cm]{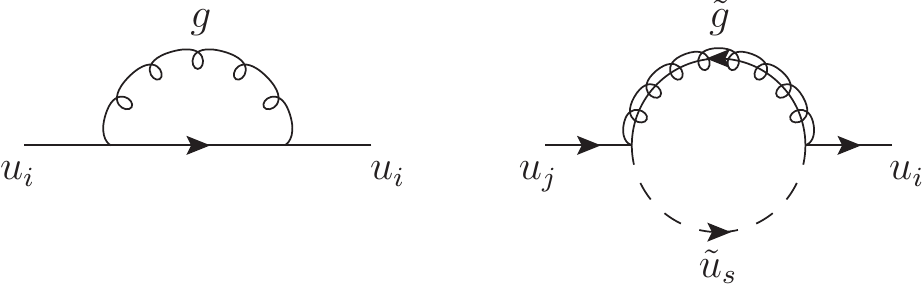}
\caption{\label{fig:virtualcorr} Generic diagrams of the
  vertex corrections (upper) and of the squark (middle) and quark
  self-energies (lower) contributing to the SUSY-QCD corrections of the decay
  $\tilde{u}_1 \to c \tilde{\chi}_1^0$, with the quark indices
  $i,j=1,2,3$ and the squark indices $r,s,t=1,...,6$.} 
\end{center}
\end{figure}

\begin{figure}[b]
\begin{center}
\includegraphics[width=15cm]{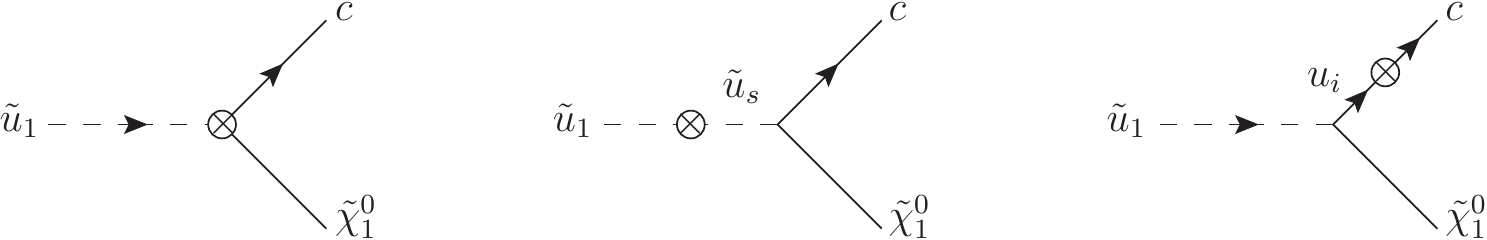}
\caption{\label{fig:counter} Counterterm diagrams.} 
\end{center}
\end{figure}

\begin{figure}[t]
\begin{center}
\includegraphics[width=8cm]{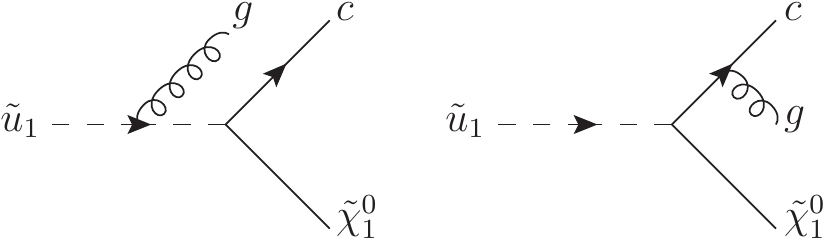}
\caption{\label{fig:real} Diagrams contributing to the real corrections.} 
\end{center}
\end{figure}

The virtual corrections
arise from the vertex diagrams shown in Fig.~\ref{fig:virtualcorr} (upper) and the
squark and quark self-energies, depicted in Fig.~\ref{fig:virtualcorr} (middle)
and (lower), respectively. The vertex corrections involve
gluons and gluinos. The gluinos can in general
couple to two different flavours of quarks and squarks, which is
taken into account by the quark and squark indices $i$ and $s$
($i=1,2,3$, $s=1,...,6$) in the corresponding second Feynman
diagram.
The counterterm diagrams in Fig.~\ref{fig:counter} cancel the
ultraviolet (UV) divergences of the virtual corrections in the
renormalisation procedure.
After renormalisation the virtual corrections still exhibit infrared
(IR) and collinear divergences. The real corrections, shown in
Fig.~\ref{fig:real}, arise from the radiation of a gluon off the squark
and off the charm quark line. In accordance with the Kinoshita-Lee-Nauenberg
theorem \cite{kln} the IR divergences emerging from the real corrections
cancel those of the virtual corrections. As there are no massless
particles in the initial state, 
in our case also the collinear divergences of the virtual and real
corrections cancel. The computation of the decay width is performed in
$D=4-2\epsilon$ dimensions. The UV and IR divergences arise then as poles in 
$\epsilon$. We distinguish between the UV and IR divergences by
denoting the corresponding poles as $1/\epsilon_{\text{UV}}$ and
$1/\epsilon_{\text{IR}}$. The loop integrals are evaluated in the
framework of dimensional reduction \cite{dimred} in order to ensure
the conservation of the SUSY relations. \s

The virtual corrections have been calculated with {\tt
  FeynArts/FormCalc}
\cite{fartsfcalc1,fartsfcalc2,fartsfcalc3,fartsfcalc4}. The results
for the gluon contribution are the same as for the squark decay into a 
quark and a neutralino, $\tilde{q}_{1,2} \to q \tilde{\chi}_1^0$, given in
Refs.\cite{stloopdecay1,stloopdecay2,stloopdecay3,stloopdecay4}. In order to
regularise the UV divergences we adopt an on-shell renormalisation scheme. The 
bare quark and squark fields with superscript $(0)$ are replaced by
the corresponding renormalised fields according to
\beq
\tilde{q}^{(0)} = \left( 1 + \frac{1}{2} \delta Z^{\tilde{q}} \right)
\tilde{q} \;, \quad\quad
q_{L,R}^{(0)} = \left( 1 + \frac{1}{2} \delta Z^{L/R} \right) q_{L/R} \;.
\eeq
In terms of the real parts of the squark self-energy $\tilde{\Sigma}$, the
squark wave function renormalisation constants $\delta Z^{\tilde{q}}$ are given by
($s,t=1,...,6$) 
\beq
\delta Z^{\tilde{q}}_{st} = \left\{ \begin{array}{ll}
- \mbox{Re} \left. \frac{\partial \tilde{\Sigma}_{ss}
  (p^2)}{\partial p^2}\right|_{p^2=m_{\tilde{q}_s}^2} & \quad \mbox{if
} s=t  \\[0.2cm] 
\frac{2}{m_{\tilde{q}_s}^2-m_{\tilde{q}_t}^2} \mbox{Re} \tilde{\Sigma}_{st}
(p^2=m_{\tilde{q}_t^2}) & \quad \mbox{if } s\ne t \;.  
\end{array}
\right.
\label{eq:deltazqtilde}
\eeq
The self-energies $\tilde{\Sigma}$ are obtained from the Feynman
diagrams in Fig.~\ref{fig:virtualcorr} (middle). Here, the third
diagram comprises the quartic squark coupling.  
As we calculate only the ${\cal O}(\alpha_s)$ corrections, in the
quartic squark coupling consistently only the terms proportional to $\alpha_s$ are
taken into account. Another diagram, not shown in
Fig.~\ref{fig:virtualcorr} (middle), involving a quartic coupling between up-
and down-type squarks, vanishes due to the flavour structure. \s

Defining the following structure for the quark self-energies ($i,j=1,2,3$),
\beq
\Sigma_{ij} (p^2) = \slash{\!\!\!p} \Sigma^L_{ij} (p^2) {\cal P}_L +
\slash{\!\!\!p} \Sigma_{ij}^R (p^2) {\cal P}_R + \Sigma^{Ls}_{ij} (p^2)
{\cal P}_L + \Sigma^{Rs}_{ij} (p^2) {\cal P}_R 
\eeq
with ${\cal P}_{L/R} = (1\mp \gamma_5)/2$, the off-diagonal chiral
components of the wave function renormalisation constants for the quarks read
\beq
\hspace*{-0.1cm}\delta Z^L_{ij} \!\!\!&=&\hspace*{-0.2cm} \frac{2}{m_{q_i}^2-m_{q_j}^2} \left[ m_{q_i} \,
  \mbox{Re} \Sigma_{ij}^{Ls} (m_{q_j}^2) + m_{q_j} \,
  \mbox{Re} \Sigma_{ij}^{Rs} (m_{q_j}^2) + m_{q_j}^2 \,
  \mbox{Re} \Sigma_{ij}^{L} (m_{q_j}^2) + m_{q_i} m_{q_j} \, 
  \mbox{Re} \Sigma_{ij}^{R} (m_{q_j}^2) \right] \nonumber \\
\hspace*{-0.1cm}\delta Z^R_{ij} \!\!\!&=&\hspace*{-0.2cm}
\frac{2}{m_{q_i}^2-m_{q_j}^2} \left[ m_{q_j} \,
  \mbox{Re} \Sigma_{ij}^{Ls} (m_{q_j}^2) + m_{q_i} \,
  \mbox{Re} \Sigma_{ij}^{Rs} (m_{q_j}^2) + m_{q_i} m_{q_j} \,
  \mbox{Re} \Sigma_{ij}^{L} (m_{q_j}^2) + m_{q_j}^2 \,
  \mbox{Re} \Sigma_{ij}^{R} (m_{q_j}^2) \right] \nonumber \\[0.1cm]
\hspace*{-0.1cm}&&\hspace*{-0.2cm}
 \mbox{for } i\ne j \; . \label{eq:deltazq} \label{eq:quarkrenconst}
\eeq
The diagonal components read ($i=j$)
\beq
\delta Z_{ii}^{L/R} &=& \hspace*{-0.2cm} - \mbox{Re} \Sigma_{ii}^{L/R} (m_{q_i}^2)
\nonumber \\
&-& \hspace*{-0.2cm} m_{q_i} \frac{\partial}{\partial p^2} \mbox{Re} \left(
  m_{q_i} \Sigma_{ii}^{L/R} (p^2) + m_{q_i} \Sigma_{ii}^{R/L} (p^2) +
  \Sigma_{ii}^{L/Rs} (p^2) + \Sigma_{ii}^{R/Ls} (p^2)
\right)\left|_{p^2=m_{q_i^2}} \right. \,.
\eeq
The self-energies $\Sigma$ appearing in the wave function
renormalisation constants are obtained from the diagrams in
Fig.~\ref{fig:virtualcorr} (lower). The gluon diagram does not contain
any scale and naively would be expected to be zero. However, it exhibits UV and IR
divergences. The diagram is proportional to $1/\epsilon_{\text{UV}} -
1/\epsilon_{\text{IR}}$ and has to be taken into account, in order
to ensure the separate cancellation of the UV and IR divergences. \s 

After the on-shell renormalisation of the quark and squark wave
functions we are only left with the one-loop vertex diagrams and the
FCNC vertex counterterm. It is given by the wave function
renormalisation, the renormalisation of the quark and squark mixing
matrices
\cite{Sirlin:1974ni,Denner:1990yz,Kniehl:1996bd,Gambino:1998ec,Yamada:2001px}
and the renormalisation of the quark masses. 
The mixing matrix counterterms
$\delta u$ and $\delta \tilde{w}$ relate the bare mixing matrices
$U^{(0)}$ and $\widetilde{W}^{(0)}$ with the renormalised ones,
\beq
\begin{array}{lcll}
U_{ij}^{(0)L/R} &=& (\delta_{ik}+\delta u_{ik}^{L/R}) U_{kj}^{L/R} \;
& \qquad i,j,k=1,2,3 \\
\widetilde{W}^{(0)}_{st} &=& (\delta_{sr}+\delta \tilde{w}_{sr}) \widetilde{W}_{rt} \;
& \qquad r,s,t=1,...,6 \;.
\end{array}
\eeq
Both the bare and the renormalised mixing matrices are required to be
unitary leading to antihermitian counterterms. We determine
the UV divergent part of each counterterm such that it cancels the
divergent part of the antihermitian part of the corresponding wave
function renormalisation matrix
\cite{Denner:1990yz,Kniehl:1996bd,Gambino:1998ec,Yamada:2001px}, 
\beq
\delta u^{L/R} &=& \frac{1}{4} \left( \delta Z^{L/R} - \delta
  Z^{L/R\dagger} \right) \\
\delta \tilde{w} &=& \frac{1}{4} \left( \delta Z^{\tilde{q}} - \delta
  Z^{\tilde{q}\dagger} \right) \;.
\eeq
The counterterms are defined on-shell. This definition of the
counterterms is known to be gauge dependent 
\cite{Gambino:1998ec,Yamada:2001px,Barroso:2000is,Kniehl:2000rb}. In
\cite{Yamada:2001px} it was stated, however, that the Feynman-'t Hooft
gauge, which we adopt here, leads to a result which coincides with the
gauge independent result. In the Yukawa
part of the squark-quark-neutralino coupling given in terms of the
left- and right-chiral couplings in Eqs.~(\ref{eq:leftchiral}) and
(\ref{eq:rightchiral}), the bare quark mass $m^{(0)}_{u_i}$ needs to be
renormalised ($i=1,2,3$), 
\beq
m^{(0)}_{u_i} = m_{u_i} + \delta m_{u_i} \;,
\eeq
with the counterterm $\delta m_{u_i}$ given by 
\beq
\delta m_{u_i} = \frac{1}{2} \mbox{Re} \left[ m_{u_i} \left(
    \Sigma^L_{ii} (m_{u_i}^2) + \Sigma^R_{ii} (m_{u_i}^2) \right) +
  \Sigma_{ii}^{Ls} (m_{u_i}^2) + \Sigma^{Rs}_{ii} (m_{u_i}^2) \right] \;. 
\eeq
Even in case of a vanishing fermion mass, a mass counterterm is
generated due to the $\Sigma^{Ls/Rs}$ contributions from the gluino
diagram in Fig.~\ref{fig:virtualcorr} (lower), see {\it e.g.}~also
\cite{Crivellin:2011sj}. In the basis of the mass eigenstates of the
squark, quark and neutralino the Lagrangian ${\cal L}_{\bar{u} \tilde{u}
  \tilde{\chi}_1^0}$ containing the vertex counterterm is then given
by
\beq
{\cal L}_{\bar{u} \tilde{u} \tilde{\chi}_1^0} = \bar{u}_i (g^L_{isl} + \delta g^L_{isl}
  ) {\cal P}_L \tilde{u}_s \tilde{\chi}_l^0 + \bar{u}_i (g^R_{isl} + \delta g^R_{isl}
  ) {\cal P}_R \tilde{u}_s \tilde{\chi}_l^0 \;,
\eeq
with the left- and right-chiral coupling counterterms ($i,j,k=1,2,3$,
$s,t=1,...,6$, $l=1,...,4$)\footnote{For a detailed derivation, see
  \cite{Muhlleitner:2011ww}.} 
\beq
\delta g^L_{isl} &=& -g e_{Rl}^{u_i} \left[ \frac{\delta
    Z^{R\dagger}_{ij}}{2} W^{\dagger}_{j+3 \, s} + \delta u^R_{ij}
  W^\dagger_{j+3 \, s} +
W^\dagger_{i+3 \, t} \delta \tilde{w}^\dagger_{ts} +
W_{i+3 \, t}^\dagger \frac{\delta Z^{\tilde{u}}_{ts}}{2} \right]
\nonumber \\
&&  -\frac{g Z_{l4}}{\sqrt{2} M_W \sin\beta} \left[
\delta m_{u_i} \delta_{ij} W^\dagger_{js} + \frac{\delta
  Z_{ij}^{R\dagger}}{2} m_{u_j} \delta_{jk} W^\dagger_{ks} + m_{u_i}
\delta u^L_{ij} W^\dagger_{js} \right. \label{eq:deltagl}\\
&& \left. + m_{u_i} \delta_{ij} W^\dagger_{jt} \delta \tilde{w}^\dagger_{ts}
+ m_{u_i} \delta_{ij} W^\dagger_{jt} \frac{\delta Z^{\tilde{u}}_{ts}}{2}\right]
\nonumber
\\
\delta g^{R}_{isl} &=& - ge_{Ll}^{u_i} \left[ \frac{\delta
    Z^{L \dagger}_{ij}}{2} W^\dagger_{js} + \delta u^L_{ij}
  W^\dagger_{js} + W^\dagger_{it} \delta \tilde{w}^\dagger_{ts} +
  W^\dagger_{it} \frac{\delta Z^{\tilde{u}}_{ts}}{2} \right] \nonumber \\
&& - \frac{g Z_{l4}}{\sqrt{2} M_W \sin\beta} \left[ \delta m_{u_i}
  \delta_{ij} W^\dagger_{j+3 \, s} + \frac{\delta Z^{L \dagger}_{ij}}{2}
  m_{u_j} \delta_{jk} W^\dagger_{k+3 \, s} + m_{u_i} \delta u^R_{ik}
  W^\dagger_{k+3 \, s} \right. \label{eq:deltagr}\\
&& \left. + m_{u_i} \delta_{ij} W^\dagger_{j+3 \, t}
  \delta \tilde{w}^\dagger_{ts} + m_{u_i} \delta_{ij} W^\dagger_{j+3
    \, t} \frac{\delta Z^{\tilde{u}}_{ts}}{2} \right] \;.
\nonumber
\eeq
Note that the wave function renormalisation constants  $\delta Z^{\tilde{u}}_{st}$
in Eq.~(\ref{eq:deltazqtilde}), and $\delta Z^{L/R}_{ij}$ in
Eq.~(\ref{eq:deltazq}) have a vanishing denominator in case of equal
masses for quarks $i$ and $j$ and squarks $s$ and $t$. This
in particular turns out to be a problem when the first and second
generation quark masses are set to zero. If both the fields and mixing
matrices are renormalised on-shell, however, this problem does not occur,
as the combination of the renormalisation constants is non-singular,
see {\it e.g.}~Ref.~\cite{Eberl:1999he}. For degenerate fermions we
hence make the replacement ($i,j=1,2,3$)
\beq
\frac{\delta Z_{ij}^{L/R \dagger}}{2} + \delta u_{ij}^{L/R} &=&
\frac{1}{4} \left( \delta Z_{ij}^{L/R} + \delta Z_{ij}^{L/R\dagger}
\right) \stackrel{m_{u_i}= m_{u_j}}{\longrightarrow} - \frac{1}{2} \left[
\mbox{Re} \Sigma^{L/R}_{ij} (m_{u_i}^2) \right. \nonumber \\
&& \hspace*{5.2cm} - m_{u_i} \frac{\partial}{\partial p^2} \mbox{Re} \left( m_{u_i}
  \left( \Sigma^{L/R}_{ij} (p^2) + \Sigma^{R/L}_{ij} (p^2) \right)
\right.  \nonumber \\
&& \left. \left. \left. \hspace*{5.2cm} + \Sigma^{L/Rs}_{ij} (p^2) +
    \Sigma^{R/Ls}_{ij} (p^2) 
\right)\right]\right|_{p^2=m_{u_i}^2} \label{eq:repl1}
\eeq
and
\beq
&& \frac{1}{2} \delta Z_{ij}^{L/R\dagger} m_{u_j} + \delta u_{ij}^{R/L}
m_{u_i} =
\frac{1}{2} \delta Z_{ij}^{L/R\dagger} m_{u_j} + \frac{1}{4} (\delta
Z^{R/L}_{ij} - \delta Z_{ij}^{R/L\dagger}) m_{u_i} \nonumber \\
&& \stackrel{m_{u_i}= m_{u_j}}{\longrightarrow} \frac{1}{2} \mbox{Re}
\left[ m_{u_i} \Sigma^{R/L}_{ij} (m_{u_i}^2) + 2 \Sigma^{R/Ls}
  (m_{u_i}^2) \right] - \frac{1}{2} \frac{\partial}{\partial p^2}
\mbox{Re} \left[ m_{u_i}^3 \Sigma_{ij}^{L/R} +m_{u_i}^3 \Sigma^{R/L}
  (p^2) \right. \nonumber \\
&& \left. \left. \phantom{\stackrel{m_{u_i}= m_{u_j}}{\longrightarrow}} + m_{u_i}^2
\Sigma^{L/Rs}_{ij} (p^2) + m_{u_i}^2 \Sigma^{R/Ls}_{ij} (p^2) \right]
\right|_{p^2=m_{u_i}^2} \;. \label{eq:repl2}
\eeq
In Eqs.~(\ref{eq:repl1})  and (\ref{eq:repl2}) we do not
sum over common indices. In the derivation of these equations we have
used 
\beq
\delta Z_{ij}^\dagger = \delta Z_{ij} (m_{u_i}^2 \leftrightarrow
m_{u_j}^2)
\eeq
in Eqs.~(\ref{eq:quarkrenconst}). This relation follows
from the hermiticity of the Lagrangian, implying that the
self-energies obey $\Sigma_{ij} = \gamma_0 \Sigma_{ij}^\dagger
\gamma_0$. For degenerate squark masses we use ($s,t=1,...,6$)
\beq
\frac{1}{2} \delta Z^{\tilde{u}}_{st} + \delta \tilde{w}_{ts}^\dagger
= \frac{1}{4} \left( \delta Z^{\tilde{u}}_{st} + \delta
  Z^{\tilde{u}\dagger}_{st} \right) \stackrel{m_{\tilde{u}_s}=
  m_{\tilde{u}_t}}{\longrightarrow} - \left. \frac{1}{2} \mbox{Re}
\frac{\partial}{\partial p^2} \tilde{\Sigma}_{st}
(p^2)\right|_{p^2=m_{\tilde{u}_s^2}} \;. 
\eeq

The real corrections have been evaluated in $D=4-2\epsilon$ dimensions
with $\epsilon \equiv \epsilon_{\text{IR}}$. As we are only interested
in the total decay width, the $D$-dimensional phase space integration
can be performed analytically. We checked explicitly that the IR and
collinear divergences of the real corrections cancel those of the
virtual corrections. Analogously we have performed 
the computation of the decay width $\Gamma(\tilde{u}_1 \to u
\tilde{\chi}_1^0)$ at NLO SUSY-QCD. All computations presented
here have been performed in two independent calculations and have been
cross-checked against each other. In Appendix A we give the explicit
formulae for the full result of the partial decay width at NLO SUSY-QCD.  \s

\section{\label{sec:4body} The Four-Body Decay}
\begin{figure}[t]
\begin{center}
\includegraphics[width=13cm]{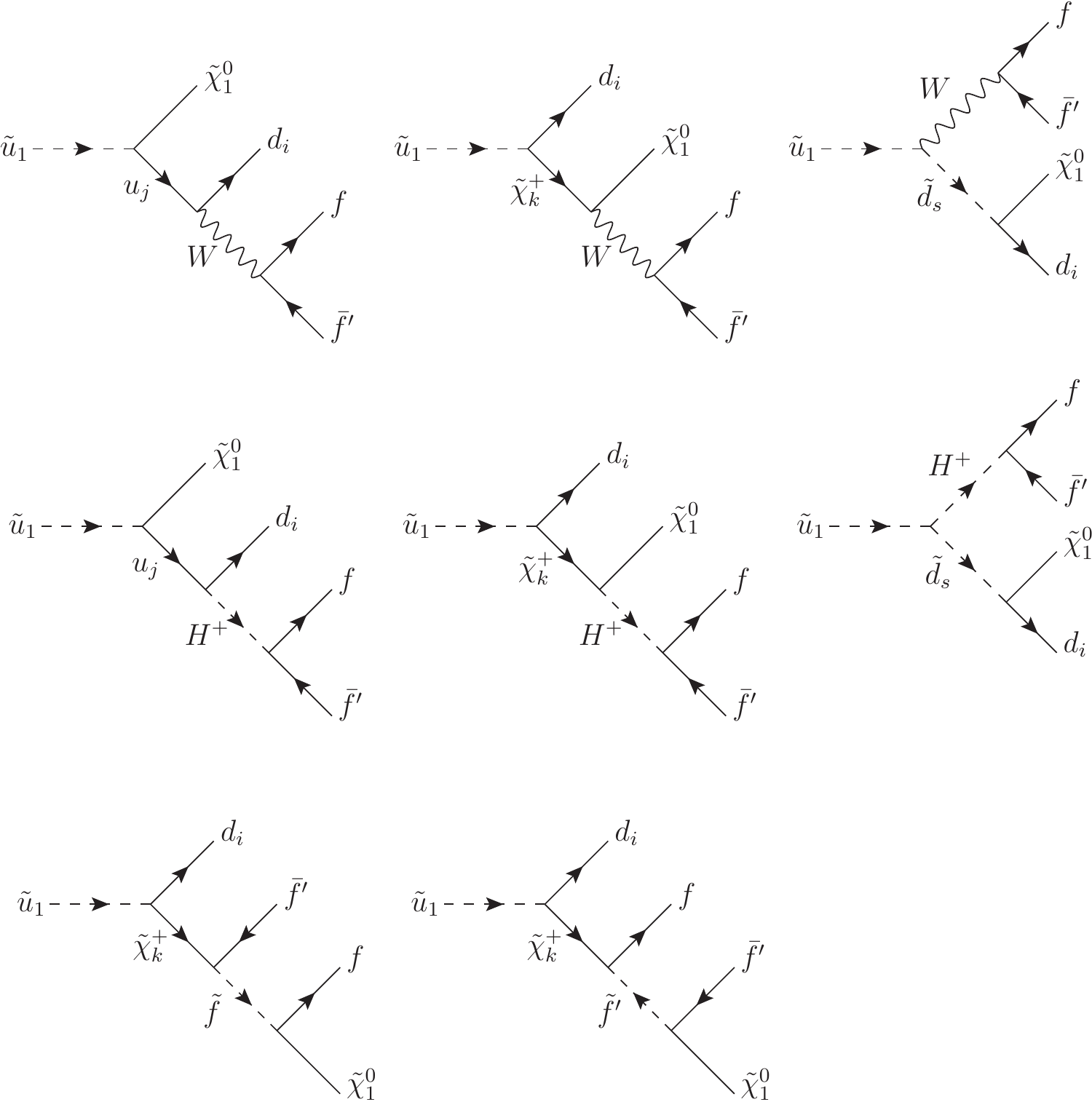}
\caption{\label{fig:4body} Generic Feynman diagrams contributing to
  the four-body decay $\tilde{u}_1 \to \tilde{\chi}_1^0 d_i f\bar{f}'$
  ($i,j=1,2,3$, $s=1,...,6$, $k=1,2$).} 
\end{center}
\end{figure}

In the parameter region where the FV two-body decay of the
lightest squark plays a role, the four-body decay into the lightest
neutralino, a down-type quark and a fermion pair can become competitive and
even dominate. The latter is in particular the case for a small FV coupling
$\tilde{u}_1-c-\tilde{\chi}_1^0$, as the four-body decay contains
flavour-conserving subprocesses. We revisit this decay, which has been
first calculated in \cite{Boehm:1999tr}, by allowing for FV couplings
at tree-level and by taking into account the 
full dependence on the masses of third generation fermions. Because of
possible flavour violation the four-body decay that we consider is
given by 
\beq
\tilde{u}_1 \to \tilde{\chi}_1^0 d_i  f\bar{f}' \;,  \label{eq:fl4bod}
\eeq
where $d_i$
denotes a down-type quark of any of the three flavours, $i=1,2,3$. The
final state fermions are $f,f'=u,d,c,s,b,e,\mu,\tau,\nu_e,
\nu_\mu, \nu_\tau$. Figure~\ref{fig:4body} 
shows the Feynman graphs contributing to the process. They are
mediated by charged Higgs $H^\pm$, $W$, chargino
$\tilde{\chi}_{1,2}^\pm$, quark and sfermion exchanges. With the
exchanged particles being far off-shell we do not take into account
total widths in the propagators, except for the one of the $W$
boson. Additionally, there are diagrams in 
which neutral particles as {\it e.g.}~neutralinos or gluinos are
exchanged and which can only proceed via FV couplings. 
These will not be considered in the numerical analysis. They are
negligibly small, as we checked explicitly. The diagrams displayed in
Fig.~\ref{fig:4body} contain fermion number flow violating
interactions, which were treated following the recipe given in
Ref.~\cite{Denner:1992vza}. The calculation of the process has been
performed in two independent approaches. One calculation was done 
automatically by using {\tt FeynArts/FormCalc}
\cite{fartsfcalc1,fartsfcalc2,fartsfcalc3,fartsfcalc4}. The second
calculation only used {\tt FeynCalc} \cite{Mertig:1990an}
to evaluate the traces. Both results were cross-checked against each
other. \s

Note that there is one caveat in the numerical evaluation of the
process. The masses of the particles, which we take from a spectrum
calculator, can potentially get quite important loop corrections. The
loop-corrected masses lead, when inserted in the propagators of the exchanged
particles, to an artificial gauge dependence of the process due to a
mismatch of the perturbative order of the particle masses and the one
of the involved couplings. We have therefore cancelled the gauge
dependence, obtained in a general $R_\xi$ gauge, before setting the
masses to their loop-corrected values. The thus obtained result
corresponds to the one given in the unitary gauge. The formulae of the
final result are quite cumbersome and lengthy so that they are not
displayed explicitly here. \s

The FV two-body and four-body decays have been
implemented in {\tt SDECAY} \cite{sdecay}, which is part of the program
package {\tt SUSY-HIT} \cite{susyhit}. Together with some follow-up
routines, for the former a new routine called {\tt SD\_lightstop2bod}
and for the latter a routine named {\tt   SD\_lightstop4bod} have been
implemented. The original version of {\tt SUSY-HIT} features the SUSY Les
Houches Accord (SLHA) \cite{slha}. As in the case of flavour
violation the SUSY Les Houches Accord 2 (SLHA2) \cite{slha2} needs to be read in,
the read-in subroutine has been modified accordingly. 

\section{\label{sec:parameterscan} The Parameter Scan}
For the numerical analysis a scan was performed in the MSSM parameter
space. The value of $\tan\beta$ and the mass of
the pseudoscalar Higgs boson $M_A$ have been varied in the ranges
\beq
1 \le \tan\beta \le 15 \qquad \mbox{and} \qquad 150 \mbox{ GeV}  \le M_A
\le 1 \mbox{ TeV }  \;.
\eeq
In our scenarios, larger values of $\tan\beta$ are disfavoured due to $B$-physics
observables. 
At tree-level $M_A$ and $\tan\beta$ determine the MSSM Higgs sector,
consisting of two 
neutral CP-even Higgs bosons $h$ and $H$, the pseudoscalar $A$ and
two charged Higgs bosons $H^\pm$. In order to shift the SM-like
neutral Higgs boson mass, given in our scenarios by the lighter scalar $h$,
to about 125~GeV, as reported by the LHC experiments \cite{higgsmass},
radiative corrections have to be taken into account, which are dominated by the
contributions from the (s)top sector. This and the determination of
the entire SUSY spectrum requires the definition of the soft SUSY
breaking masses and trilinear couplings. 
The Higgs and SUSY spectra have been
obtained from the spectrum calculator {\tt SPheno} \cite{spheno},
which allows for flavour violation.\footnote{We cross-checked the
  results against {\tt 
    SOFTSUSY} \cite{Allanach:2001kg}. In general the results agree
  well. However, in particular for low mass values of the lightest
  squark, there can be differences in the mass values and in the 
  squark mixing matrix elements. They are due to a different treatment of 
loop corrections in the squark mass matrices.}  The program package
{\tt SPheno} reads in the parameters 
in the SLHA2. In the SLHA format all input parameters listed here
below are understood as $\overline{\mbox{DR}}$ parameters given at the scale
$M_{\text{input}}$.\footnote{The only exception is $\tan\beta$ which is
defined as $\overline{\text{DR}}$ parameter at the scale of the $Z$
boson mass $M_Z$.} After the application of renormalisation group running
the parameters, masses and mixing values are given out in the SLHA
format at a user defined output scale 
$M_{\text{output}}$.  We chose both the input and the output scale as 
\beq
M_{\text{input}} = M_{\text{output}} = 300 \mbox{ GeV} \;.
\eeq
This is within the mass range of the lightest stop resulting from our
parameter scan. The input soft SUSY breaking gaugino mass parameters
have been chosen as 
\beq
75 \mbox{ GeV} \le M_1 \le 500 \mbox{ GeV} \; , \quad
M_2 = 650 \mbox{ GeV} \quad \mbox{and} \quad M_3 = 1530 \mbox{ GeV}
\;. 
\eeq
The lower bound on $M_1$ restricts neutralino masses to
values in accordance with the bounds from the relic density and the
ones resulting from light stop mass searches. The chosen value for
$M_3$ leads to heavy enough gluino masses to avoid the LHC exclusion
bounds. The higgsino mass parameter has been set to 
\beq
\mu = 900 \mbox{ GeV} \;.
\eeq
The obtained chargino masses are of the order of several hundred GeV
and not in conflict with any exclusion bounds. 
The soft SUSY breaking trilinear couplings and mass parameters of the
slepton sector, the right-handed up-type squark mass parameters and trilinear
couplings of the first and second generations, and the right-handed
down-type mass parameters and trilinear couplings of 
all three generations have been chosen as ($E \equiv e, \mu, \tau$,
$U \equiv u,c$, $D \equiv d,s,b$)
\beq
\begin{array}{lclcll}
M_{\tilde{E}_R} &=& M_{\tilde{L}_{1,2,3}} &=& 1 \mbox{ TeV} \; , & \quad A_E = 0
\mbox{ TeV} \; ,\\
M_{\tilde{U}_R} &=& M_{\tilde{D}_R} &=& 1.5 \mbox{ TeV} \; , & \quad A_U =
A_D = 0 \mbox{ TeV} \;.
\end{array}
\eeq
We do not apply strict MFV, but allow the right-handed stop mass parameter and 
the top trilinear coupling to vary in the range 
\beq
300 \mbox{ GeV} \le M_{\tilde{t}_R} \le 600 \mbox{ GeV} \quad
\mbox{and} \quad 1 \mbox{ TeV}  \le A_t \le 2 \mbox{ TeV}\;.
\eeq
Furthermore, we implemented two different flavour symmetries of the 
squark sector, a $U(2)_{Q_L}\times U(2)_{u_R} \times U(3)_{d_R}$
symmetry, to which we refer as $U(2)$ in the following, and a $U(3)_{Q_L}\times 
U(2)_{u_R} \times U(3)_{d_R}$ symmetry, to which we refer as $U(3)$,
{\it i.e.}\footnote{After application of all constraints the induced
  flavour violation turns out to be small, also in the non-MFV
  scenario that we apply in our analysis.}
\beq
\begin{array}{lcl}
U(2) &:& M_{\tilde{Q}_{1}} =  M_{\tilde{Q}_{2}}  = 1.5 \mbox{ TeV}
\quad \mbox{and} \quad 1 \mbox{ TeV} \le M_{\tilde{Q}_3} \le 1.5 \mbox{ TeV} \\[0.2cm]
U(3) &:& 1 \mbox{ TeV} \le M_{\tilde{Q}_{1}} =  M_{\tilde{Q}_{2}}  =
M_{\tilde{Q}_{3}}  \le 1.5 \mbox{ TeV} \;.
\end{array} \label{eq:flsymm}
\eeq
With these parameter values the squarks of the first two
generations are heavy enough not to be excluded by the experiments. 
The choice of the soft SUSY breaking parameters in the stop sector
guarantees rather low lightest stop mass values, which we are
interested in here. The SM input parameters as required by the SLHA 
are set to the particle data group (PDG) \cite{pdg} values
\beq
\begin{array}{rclrcl}
G_F &=& 1.166379 \cdot 10^{-5} \mbox{ GeV}^{-2}\,,  & \alpha_s
(M_Z)_{\overline{\text{MS}}} &=& 0.1185\,, \\[0.1cm]
m_b (m_b)_{\overline{\text{MS}}} &=& 4.18 \mbox{ GeV} \,,& m_t
(\text{pole}) &=& 173.07 \mbox{ GeV} \,, \\[0.1cm]
m_{\tau} (\text{pole}) &=& 1.77682 \mbox{ GeV} \,, &
M_Z (\text{pole}) &=& 91.1876 \mbox{ GeV} \,.
\end{array}
\eeq
Finally, according to the SLHA2 we need the CKM matrix elements in the Wolfenstein
parametrisation. The values given by the PDG are
\beq
\lambda = 0.22535 \;, \; A= 0.811 \;, \; \bar{\rho} = 0.131 \; , \;
\bar{\eta} = 0.345 \;.
\eeq
From the scan only those points are retained that lead to a mass
difference $\Delta m$ between the lightest squark $\tilde{u}_1$ and
the lightest neutralino $\tilde{\chi}_1^0$ of 
\beq
5 \mbox{ GeV} \le \Delta m = m_{\tilde{u}_1} - m_{\tilde{\chi}_1^0} \le 75
\mbox{ GeV}  \;, \label{eq:deltam}
\eeq
and that in addition fulfill the constraints we apply. The constraints arise
from the searches for Higgs boson(s) and SUSY particles, from the
relic density measurements and from flavour observables. In detail:
\s

\noindent
\underline{\it Constraints from Higgs data:} The compatibility with the
experimental Higgs data is checked with the programs {\tt HiggsBounds}
\cite{higgsbounds} and {\tt HiggsSignals} \cite{higgssignals}. The
program {\tt HiggsBounds} needs as inputs the effective couplings of
the Higgs bosons of the model under consideration, normalised to the
corresponding SM values, as well as the masses, the widths and the
branching ratios of the Higgs bosons. It then checks for the compatibility with the
non-observation of the SUSY Higgs bosons, in particular whether the
Higgs spectrum is excluded at the 95\% confidence 
level (CL) with respect to the Tevatron and LHC measurements or
not. The package {\tt HiggsSignals} on the other hand takes the same
input and validates the compatibility of the SM-like Higgs boson with the
data from the observation of a Higgs boson. As result a $p$-value is
given out, which we demanded to be at least 0.05, corresponding to a
non-exclusion at 95\% CL. For the computation of the effective
couplings and decay widths of the SM and MSSM Higgs bosons, the Fortran
code {\tt HDECAY} \cite{hdecay} is used, which provides the SM and
MSSM decay widths and branching ratios including the state-of-the-art
higher order corrections. As {\tt HDECAY} does not 
support flavour violation, the dominant flavour-diagonal entries of
the mass and mixing matrices provided by {\tt SPheno} have been
extracted before passing them on to {\tt HDECAY}. Since
FV effects in the Higgs decays are tiny and far beyond
the experimental precision, the effect of this procedure on the final
results is negligible. \s

\begin{figure}[t]
\begin{center}
\includegraphics[width=12cm]{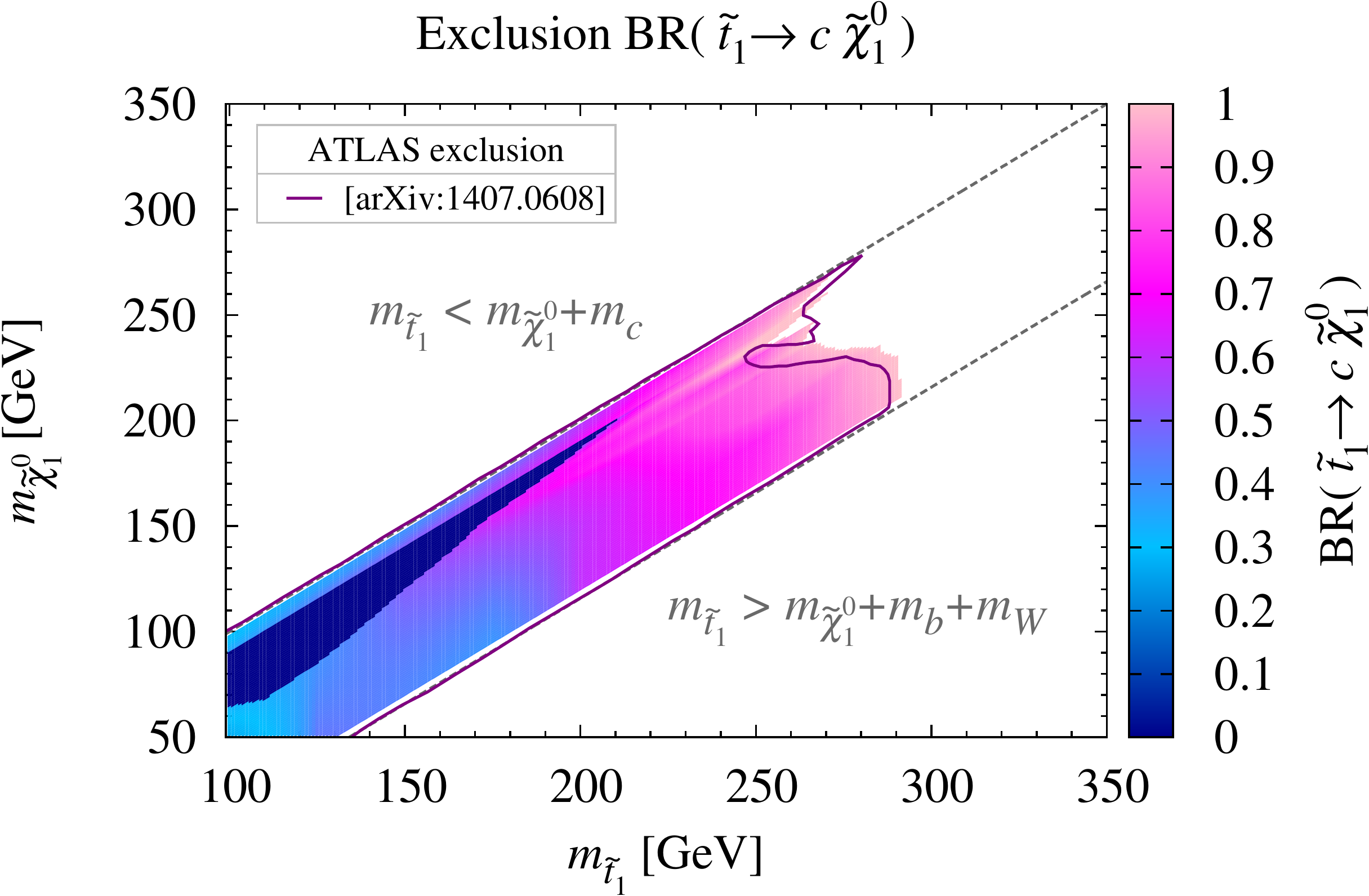} \\[0.5cm]
\includegraphics[width=12cm]{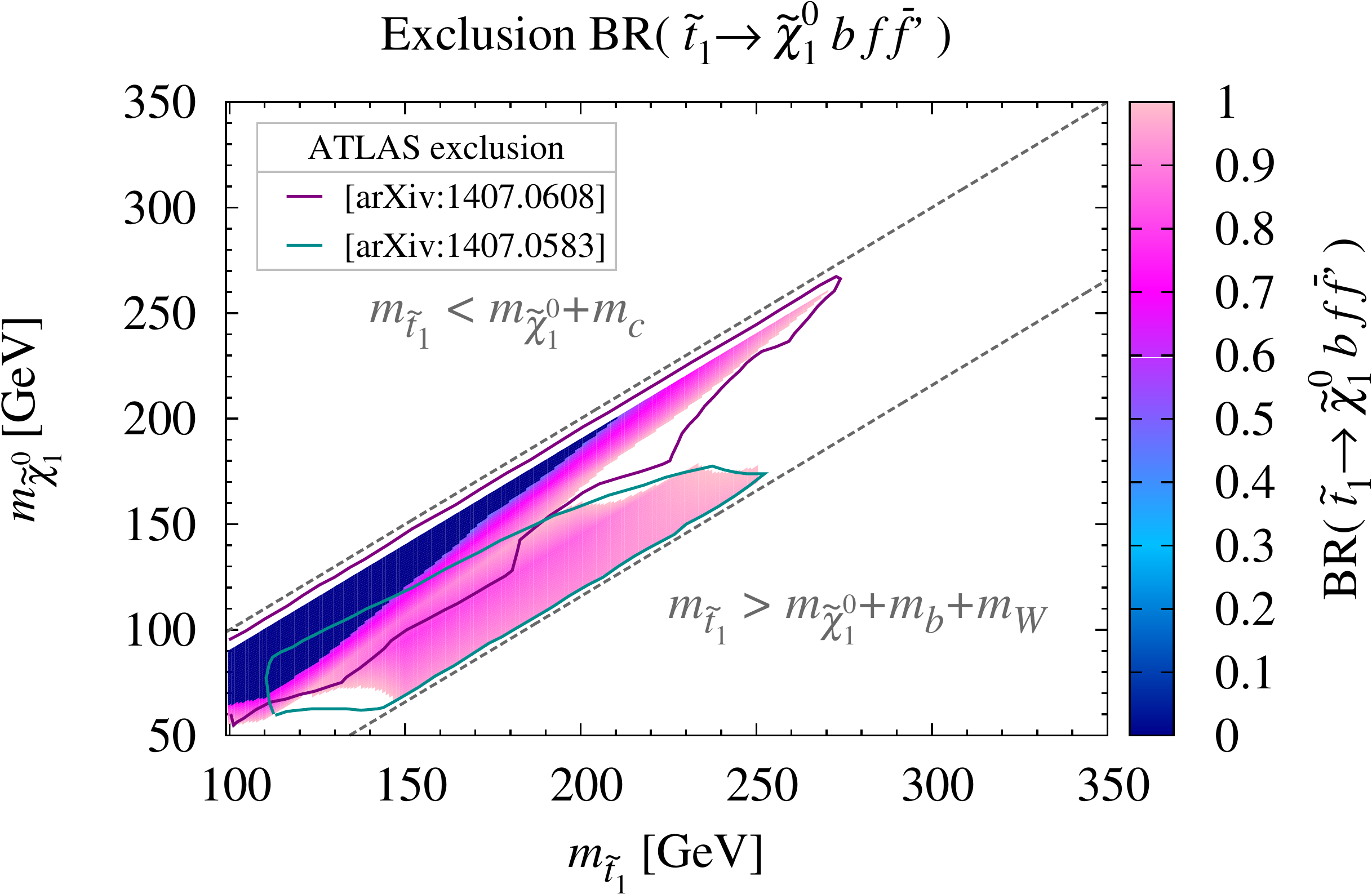} 
\caption{Exclusion limits in the
  $m_{\tilde{\chi}_1^0}-m_{\tilde{t}_1}$ plane at 95\% CL, based on
  the results for the $\tilde{t}_1 \to c \tilde{\chi}_1^0$ signature
  from \cite{atlas2} (upper) and on the results for the $\tilde{t}_1
  \to \tilde{\chi}_1^0 b f\bar{f}'$ signature from
  \cite{atlas2,atlas3} (lower).  The colour code indicates the
  branching ratio down to which the exclusion limits are
  valid. \label{fig:stopexclusion}}   
\end{center}
\end{figure}

\noindent
\underline{\it Constraints from SUSY searches:} In order not to be in
conflict with the SUSY mass bounds reported by the LHC experiments for
the gluino and squark masses of the first two
generations \cite{atlassusybound,cmssusybound}  we required these SUSY
particles to have masses of 
\beq
m_{\tilde{g}} > 1450 \mbox{ GeV} \quad \mbox{and} \quad m_{\tilde{q}_{1,2}}
> 900 \mbox{ GeV} \quad (q=u,c,d,s) \;.
\eeq
At the LHC, searches have been performed for the lightest stop
with mass close to the LSP assumed to be $\tilde{\chi}_1^0$
in the two decay channels we are
interested in here, the flavour-changing two-body decay
Eq.~(\ref{eq:fl2bod}) and the four-body decay
Eq.~(\ref{eq:fl4bod}). Based on monojet-like \cite{atlas1,atlas2,cms1}
and charm-tagged event selections \cite{atlas1,atlas2} and on searches
for final states with one isolated lepton, jets and missing transverse momentum
\cite{atlas3}, limits are given on the lightest 
stop mass as a function of the neutralino mass, assuming,
respectively, a branching ratio of one, depending on the final state under
investigation. At present, the most stringent bounds have been
reported in \cite{atlas2,atlas3} for $\tilde{t}_1$ masses down to 
$\sim 100$~GeV. Giving up the assumption of maximum branching ratios, we
re-interpreted these limits for arbitrary stop branching ratios
below one. The results are shown in Fig.~\ref{fig:stopexclusion} in the
$m_{\tilde{\chi}_1^0}-m_{\tilde{t}_1}$ plane.\footnote{To match the
  notation of the LHC experiments we here denote $\tilde{u}_1$ by
  $\tilde{t}_1$, which is approximately the case for small flavour
  violation.} The grey dashed lines limit the region in which 
\beq
m_{\tilde{\chi}_1^0} + m_c \le m_{\tilde{t}_1} \le
m_{\tilde{\chi}_1^0} + m_b + m_W \;. \label{eq:massinterval}
\eeq 
In this region the stop can be searched for in the
FV two-body decay Eq.~(\ref{eq:fl2bod}) and the four-body
decay Eq.~(\ref{eq:fl4bod}). Neglecting the two-body decay
$\tilde{t}_1 \to u \tilde{\chi}_1^0$, which is usually suppressed by two orders
of magnitude compared to the two-body decay with the charm quark final
state, the $\tilde{t}_1$ branching ratios in this mass region are given by
\beq
\mbox{BR} (\tilde{t}_1 \to c \tilde{\chi}_1^0) &=&
\frac{\Gamma(\tilde{t}_1 \to c \tilde{\chi}_1^0)}{\Gamma_{\text{tot}}} \\ 
\mbox{BR} (\tilde{t}_1 \to \tilde{\chi}_1^0 b f \bar{f}') &=&
\frac{\Gamma(\tilde{t}_1 \to \tilde{\chi}_1^0 b f
  \bar{f}')}{\Gamma_{\text{tot}}} \;, \qquad \qquad \qquad \qquad
\mbox{with} \\ 
\Gamma_{\text{tot}} &=& \Gamma(\tilde{t}_1
  \to c \tilde{\chi}_1^0) + \Gamma(\tilde{t}_1 \to \tilde{\chi}_1^0 b
f \bar{f}') \;.
\eeq
The full pink line in the upper plot shows
the 95\% CL exclusion limit based on combined charm-tagged and monojet
ATLAS searches\footnote{The exclusion limits do not apply for the $u
  \tilde{\chi}_1^0$ final state. In principle monojet searches could
  be used to derive limits in this decay channel.} in the $\tilde{t}_1
\to c\tilde{\chi}_1^0$ decay \cite{atlas2}, assuming 
100\% branching ratio. For a $\tilde{t}_1$ decaying exclusively into
the four-body final state ATLAS derived from the monojet analysis
\cite{atlas2} the exclusion given by the pink line (close to the upper
dashed line) in Fig.~\ref{fig:stopexclusion} (lower) and from the
final states with one isolated lepton the exclusion region delineated by the green
line (close to the lower dashed line) \cite{atlas3}. With the
information given in \cite{atlas2,atlas3} we 
derived the exclusion limits for the two- and the four-body final
state as a function of the branching ratio, which is given by the
colour code. For the exclusion limits of Ref.~\cite{atlas2} we derived
the limits with the CL$_s$ method \cite{Read:2002hq}. Uncertainties on
both background and  signal were taken into account by
Gaussian probability distribution functions. In the derivation of the
limits in the four-body final state we assumed that the branching
ratio with jet final states, $\mbox{BR} (\tilde{t}_1 \to 
\tilde{\chi}_1^0 b jj)$, makes up 66\%, and the branching ratios 
$\mbox{BR} (\tilde{t}_1 \to \tilde{\chi}_1^0 b \bar{l} \nu_l)$ ($l=e,\mu,\tau$)
each account for 11\% of the four-body decay branching ratio, see also the
discussion on the four-body decay branching ratio in
subsection~\ref{subsec:fourbody}. \s

From the plots it can be read off that stop masses with a branching
ratio above the one associated  
with a specific colour are excluded. It is immediately evident
that for smaller branching ratios the exclusion limits become
weaker. The two plots can be combined to extract the exclusion limits
for stops of a given mass as function of the neutralino mass and the
stop branching ratio. Thus it can be read off from
Fig.~\ref{fig:stopexclusion} (upper) that $\tilde{t}_1$ masses of
150~GeV can be excluded for $m_{\tilde{\chi}_1^0}=80$~GeV if their branching
ratio into $c + \tilde{\chi}_1^0$ exceeds $0.43$. This in turn
implies that the stop four-body branching ratio is below 0.57. On
the other hand the lower plot shows that in the same region stops can
be excluded if their branching ratio into the four-body final state
is larger than 0.88, which implies that the two-body decay branching
ratio is below 0.12 then. This means
that $m_{\tilde{t}_1} = 150$~GeV can be excluded for
$m_{\tilde{\chi}_1^0}=80$~GeV for scenarios in which $\mbox{BR}
(\tilde{t}_1 \to c \tilde{\chi}_1^0) < 0.12$ and $\mbox{BR}
(\tilde{t}_1 \to c \tilde{\chi}_1^0) > 0.43$, respectively, 
$\mbox{BR} (\tilde{t}_1 \to \tilde{\chi}_1^0 b f \bar{f}') >
0.88$ and $\mbox{BR} (\tilde{t}_1 \to \tilde{\chi}_1^0 b f \bar{f}') < 0.57$.
The dark blue region corresponds to stop branching ratios that are zero,
so that all stop mass values associated with this region are
excluded. In Fig.~\ref{fig:stopexclusion} (upper) there is no smooth
transition between the dark blue and its neighbouring regions, as the exclusion
limits in the two-body final state are related to the ones in the
four-body final state which here apply for branching ratios $\gsim
0.46$, so that of course also in
Fig.~\ref{fig:stopexclusion} (lower) there is no continuous colour
gradient here. \s

Our exclusion limits given by the border of the coloured
region at 100\% two-, respectively, four-body decay branching ratio, do not 
exactly match the ones derived by ATLAS. The reason is that ATLAS
provided information on the values of the excluded production cross
section times branching ratio only for a few points in the
$m_{\tilde{\chi}_1^0}-m_{\tilde{t}_1}$ plane and we had to interpolate
linearly between these points in order to cover the whole
region. Nevertheless, the agreement of our results with the given
exclusion limits is reasonably good. We take the thus derived exclusion limits
as function of the stop branching ratio in order to restrain our 
parameter points to the experimentally allowed values. The advantage
of our approach is to take fully into account the information on the
actual stop branching ratios which can considerably weaken the stop
exclusion limits as is evident from Fig.~\ref{fig:stopexclusion}.  As
our plots can only be a rough approximation of what can be
done much more accurately by the experiments, they should be taken as
an encouragement to provide results also as function of the stop
branching ratios. \s

\noindent
\underline{\it Constraints from relic density and $B$-physics
  measurements:}  
The space telescope {\tt PLANCK} \cite{Ade:2013zuv} has measured the
relic density of Dark Matter (DM) to be
\beq
\Omega_c h^2 = 0.1199 \pm 0.0027 \;.
\eeq
In our set-up we assume the lightest neutralino to be the LSP and hence 
the DM candidate. We have used the program {\tt SuperIso Relic}
\cite{superisorelic} to calculate the relic density for neutralino DM and
have compared the outcome to the experimental value. We require the
relic density resulting from neutralinos to be
\beq
\Omega_c h^2 (\tilde{\chi}_1^0) < 0.12 \;,
\eeq
which means that neutralinos are assumed not to be the only source
contributing to the measured relic density. \s

Further constraints arise from flavour observables. In particular, in
models with FCNC couplings at tree-level new particles can have a significant
impact on rare meson decays mediated by loops. We use the program
{\tt SuperIso} \cite{superiso} to calculate the relevant $B$ meson
branching ratios and require them to be compatible within two standard
deviations with the experimentally measured values. With the errors
denoting the one sigma bounds, they are given by  
\beq
\begin{array}{rcll}
{\cal B} (B_s^0 \to \mu^+ \mu^-) &=& (2.9\pm 0.7) \times 10^{-9} &
\qquad \mbox{\cite{bsubsmumu}} \\
{\cal B} (B^0 \to \mu^+ \mu^-) &<& 8.1 \times 10^{-10} \; \mbox{at }
95\% \mbox{ CL} & \qquad \mbox{\cite{b0mumu}} \\
{\cal B} (B^+ \to \tau^+ \nu_\tau) &=& (1.05 \pm 0.25) \times 10^{-4} &
\qquad \mbox{\cite{pdg}} \\
{\cal B} (B\to X_s \gamma) &=& (355 \pm 24 \pm 9) \times 10^{-4} &
\qquad \mbox{\cite{bxsgamma}} \;.
\end{array}
\eeq
We do not use the measured value of the anomalous magnetic moment
$a_\mu$ as constraint, as the SUSY contribution resulting from 
our parameter scan cannot explain the discrepancy between the SM
prediction and the experimental value. \s

For completeness we give the $\tilde{u}_1$ masses that we obtain as a result of
our scan and after application of all constraints. For the two flavour
scenarios they are 
\beq
\begin{array}{lclcccl}
U(2) & : & 223 \mbox{ GeV} &\lsim& m_{\tilde{u}_1} &\lsim& 527 \mbox{ GeV} \\
U(3) & : & 195 \mbox{ GeV} &\lsim& m_{\tilde{u}_1} &\lsim& 525 \mbox{ GeV} \;.
\end{array}
\label{eq:massvalues}
\eeq
The charged Higgs boson masses range between about 380 and 1000~GeV, the
chargino masses between approximately 655 and 665~GeV. 

\section{\label{sec:numerical} Numerical Results}
In the following we present results for the parameter points of our
scan that pass the constraints discussed in
Section~\ref{sec:parameterscan}. 
Two scenarios of flavour violation are investigated in the left-handed
squark sector, one with a  flavour symmetry $U(2)$ and a second where
the flavour symmetry is enhanced to $U(3)$, {\it
  cf.}~Eq.~(\ref{eq:flsymm}). Furthermore, the decay
$\tilde{u}_1 \to u \tilde{\chi}_1^0$ has been included in the total
width everywhere where applicable. When we talk about the FCNC decay
in the following, we implicitly refer to the $\tilde{u}_1 \to c \tilde{\chi}_1^0$ decay,
however, as the decay with the up quark final state is negligible compared to the
one with the charm quark in the final state.\s 

\subsection{SUSY-QCD Corrections to the FCNC Two-Body Decay}
\begin{figure}[h]
\begin{center}
\hspace*{-0.6cm}
\includegraphics[width=7.9cm]{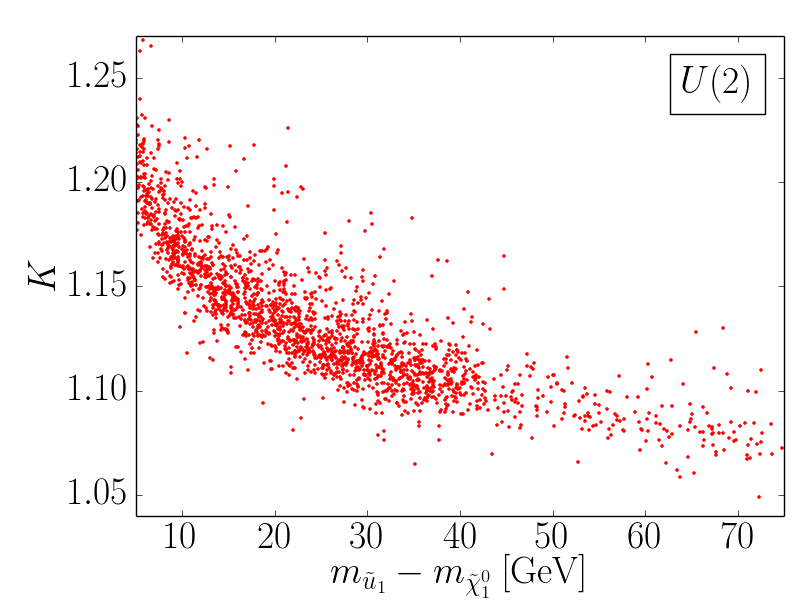} 
\includegraphics[width=7.9cm]{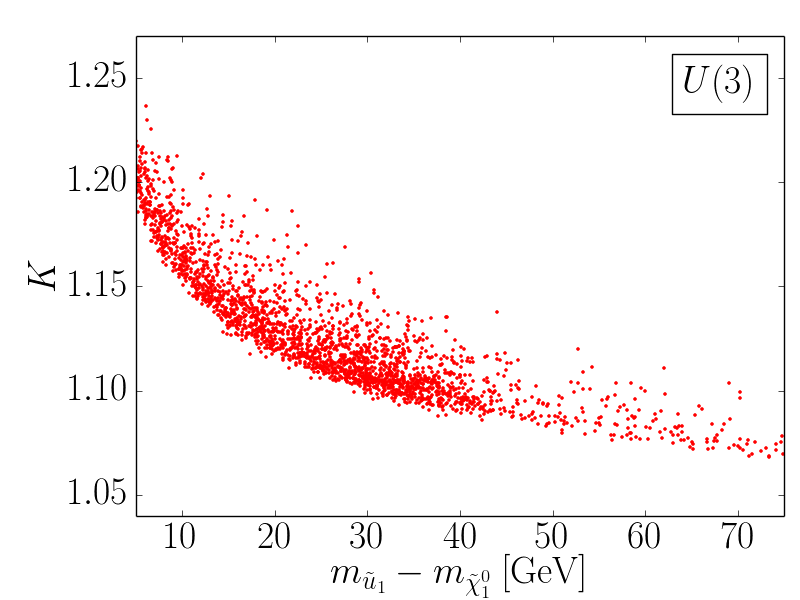} 
\caption{The SUSY-QCD $K$-factor for the FCNC decay $\tilde{u}_1 \to c
  \tilde{\chi}_1^0$ as a function of the squark-neutralino mass
  difference assuming a $U(2)$ (left) and a $U(3)$ (right) symmetry in the
  left-handed squark sector. \label{fig:kfactor}}   
\end{center}
\end{figure}

We first analyse the effect of the SUSY-QCD corrections on the
two-body decay $\tilde{u}_1 \to c \tilde{\chi}_1^0$. 
Figure~\ref{fig:kfactor} shows the $K$-factor, {\it i.e.}~the
ratio of the NLO decay width with respect to the LO decay width, as a
function of the mass difference $\Delta m = m_{\tilde{u}_1}-
m_{\tilde{\chi}_1^0}$. The strong coupling constant has been evaluated
in the $\overline{\mbox{DR}}$ scheme at the scale $m_{\tilde{u}_1}$. 
As stated in Eq.~(\ref{eq:deltam}) we vary
$\Delta m$ between 5 and 75 GeV, which on the lower and upper bounds
corresponds to the lightest squark mass interval
Eq.~(\ref{eq:massinterval}), in which the two-body (and also the
four-body) decay is relevant, modulo an off-set of a few GeV. The lower off-set
of 5~GeV accounts for the fact, that we have not taken into account
the finite charm quark mass in the two-body decay. With 5~GeV we are
far enough away from the threshold so that finite charm quark mass
effects are negligible. Note also, that for a stop mass too close to
the neutralino mass its lifetime becomes larger than the
flight time within the detector. The upper bound takes into account
that for a meaningful 
prediction in the mass region where the three-body off-shell decay
becomes important a smooth interpolation between the two- (also the four-) and
the three-body decays is required, that is not available at
present. 
As can be inferred from Fig.~\ref{fig:kfactor}, the SUSY-QCD corrections are
significant and vary between at most $\sim 27$\% ($\sim 24$\%) to
about 5\% (7\%) for $U(2)$ (for $U(3)$) when going from $\Delta m=
5$~GeV to 75~GeV. The $K$-factor increases for small mass differences, where the
real corrections become more important and increase the partial
width. The virtual corrections on the other hand decrease the partial
width, but less strongly, so that the net effect is a $\sim 27$\%
increase of the loop-corrected width for $U(2)$, respectively $\sim
24$\% for $U(3)$. For large mass differences the
$K$-factors for the real and the virtual corrections approach 1
from above and below, respectively, resulting in a residual 5-7\% correction
for the overall $K$-factor.\footnote{Note that in the branching ratios
the effect of the SUSY QCD corrections is less important, increasing
them by a few percent at NLO.} The plots also
show that for an increased 
flavour symmetry, as assumed in Fig.~\ref{fig:kfactor} (right), the
points scatter less. In the more flavour-symmetric case the
flavour off-diagonal matrix elements in the  squark mixing matrix are smaller
and the scan over the parameter space leads to smaller overall
variations of the mixing matrix elements relevant for the
$\tilde{u}_1-c-\tilde{\chi}_1^0$ coupling entering the FCNC
two-body decay. Therefore, the spread in the results for the decay widths
of the scan parameter points is less important for the $U(3)$ than for
the $U(2)$ flavour symmetry. 

\subsection{The Four-Body Decay \label{subsec:fourbody}}
The new element in our calculation of the four-body decay
$\tilde{u}_1 \to \tilde{\chi}_1^0 d_i f \bar{f}'$ compared to the
literature \cite{Boehm:1999tr} is the inclusion of FCNC couplings at
tree-level and the inclusion of the full mass dependence of the final
state bottom quarks and $\tau$ leptons. 
\begin{figure}[t]
\begin{center}
\vspace*{-0.4cm}
\hspace*{-0.4cm}
\includegraphics[width=7.9cm]{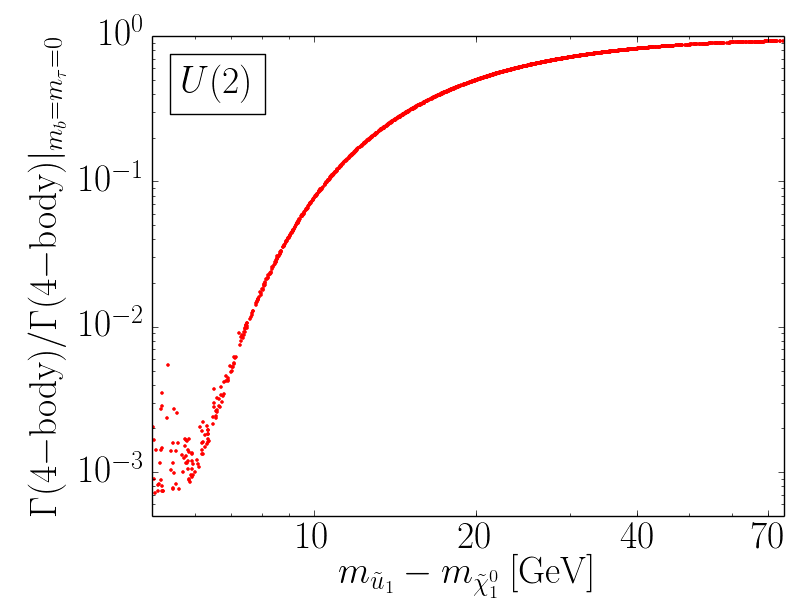} \hspace*{0.1cm}
\includegraphics[width=7.9cm]{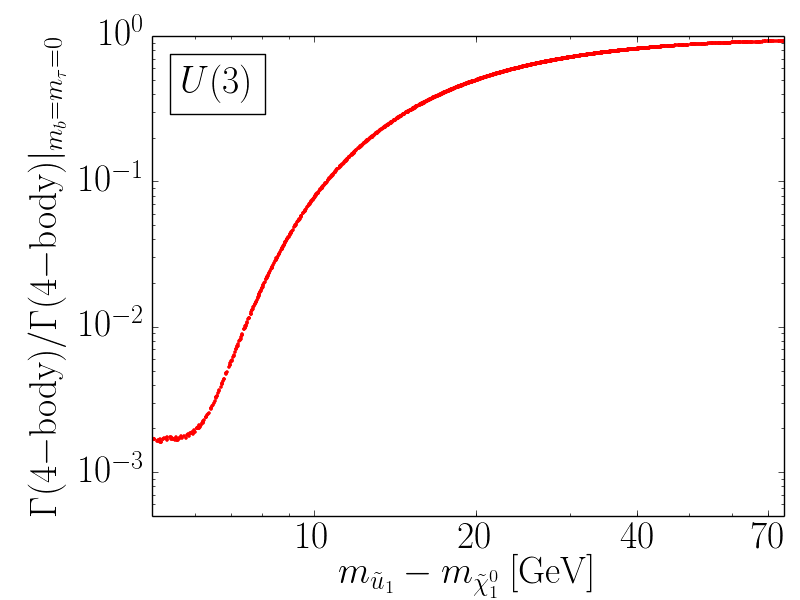} \\
\hspace*{-0.4cm}
\includegraphics[width=7.9cm]{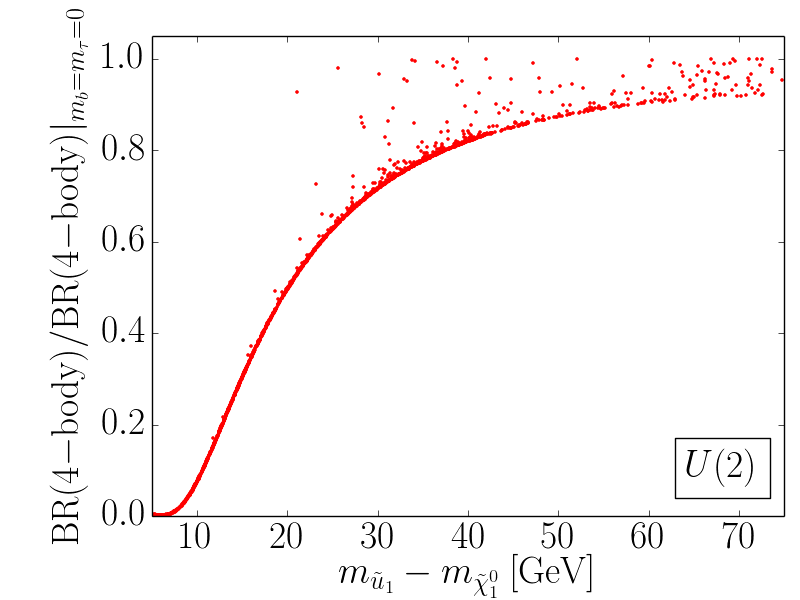} \hspace*{0.1cm}
\includegraphics[width=7.9cm]{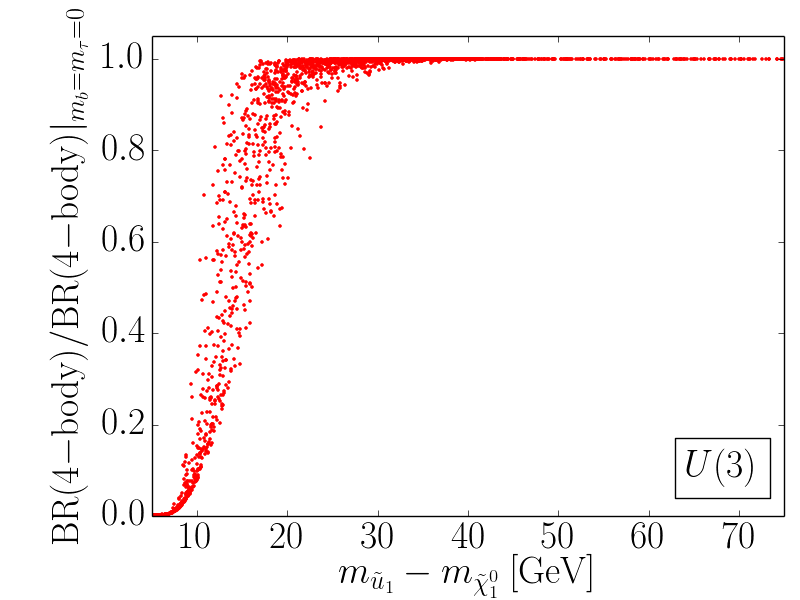} 
\caption{Upper: Ratio of the partial width for the $\tilde{u}_1 \to
  \tilde{\chi}_1^0 d_i f \bar{f}'$ ($i=1,2,3$) decay with non-zero $m_b$ and
  $m_\tau$ and of the corresponding decay width with zero
  masses. Lower: Same as upper, but for the branching ratios. 
In the left-handed squark sector a $U(2)$ (left) or a $U(3)$ (right) symmetry
is assumed. \label{fig:masseffect}}   
\end{center}
\end{figure}
The effect of taking into account non-vanishing $m_b$ and $m_\tau$ is
shown in the plots of Fig.~\ref{fig:masseffect} (upper), which show the ratio of
the partial four-body decay width with non-vanishing masses and the
corresponding width, where $m_b = m_\tau=0$, as a function of $\Delta
m$. As expected, as soon as $\Delta m$ crosses the threshold of $m_b$
the ratio steeply increases to reach an almost constant value of 
0.92 for large $\Delta m$, both for the $U(2)$ and the $U(3)$ 
symmetry. Below the threshold the ratio does not become zero due to
the diagrams contributing to the four-body decay which proceed via
FCNC couplings leading to massless final states, {\it e.g.}
$\tilde{u}_1 \to \tilde{\chi}_1^0 d_{1} \bar{e} \nu_e$. The ratio
scatters over a wider range for $U(2)$, {\it
  cf.}~Fig.~\ref{fig:masseffect} (upper left), than for $U(3)$,  {\it
  cf.}~Fig.~\ref{fig:masseffect} (upper right), due to
the lower flavour symmetry in the former case. While the mass effect
in the partial width with up to 8\% even in the region far above
threshold is non-negligible, in the 
branching ratio it gets more and more washed out with increasing
importance of the four-body decay width,
{\it cf.}~Fig.\ref{fig:masseffect} (lower). As in case of the $U(3)$
symmetry for large $\Delta m$ the four-body decay dominates over the
two-body decay, {\it cf.}~next subsection, the mass effect becomes
almost zero in the branching ratio then. For the $U(2)$ symmetry it
can still be up to 10\% for $\Delta m=75$~GeV. In the threshold region, the
mass effect in the branching ratios is important and has to be taken
into account as it is phenomenologically relevant, see also the
discussion on the comparison between two- and four-body $\tilde{u}_1$
decays below.\s 

\begin{figure}[t]
\begin{center}
\vspace*{-0.4cm}
\hspace*{-0.5cm}
\includegraphics[width=7.9cm]{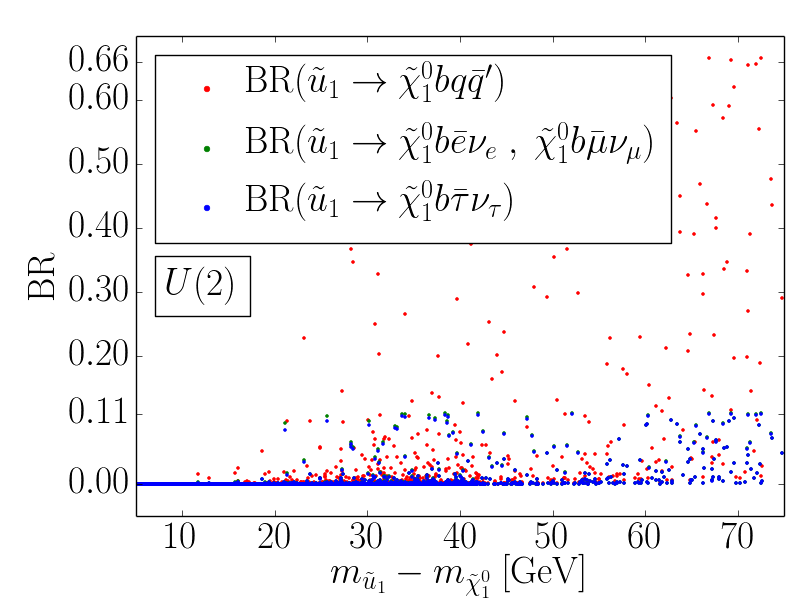} \hspace*{0.1cm}
\includegraphics[width=7.9cm]{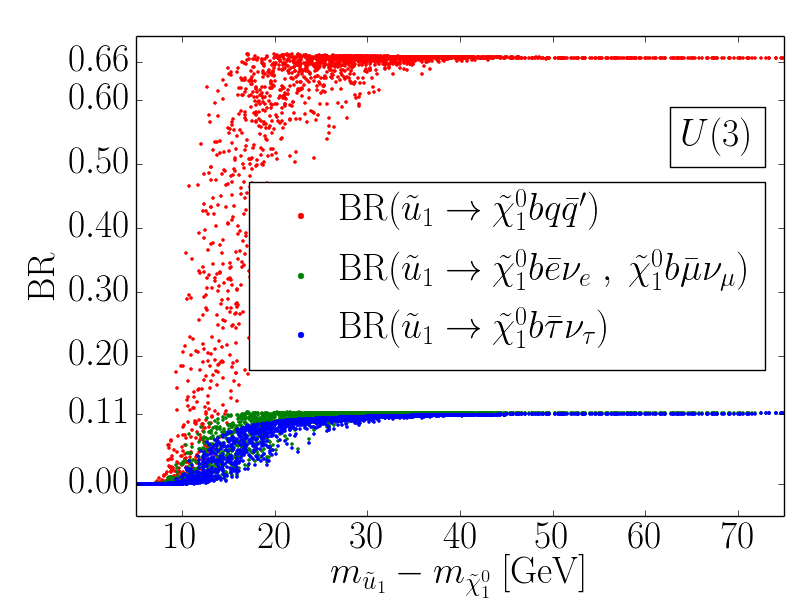} 
\caption{The dominant final state branching ratios of the four-body
  decay, assuming in the left-handed squark sector a $U(2)$ (left) or a $U(3)$ (right) symmetry. \label{fig:4bodydominant}}   
\end{center}
\end{figure}
In Fig.~\ref{fig:4bodydominant} the branching ratios of the dominant final state
signatures to the four-body decay are shown, {\it i.e.} $\tilde{u}_1 \to
\tilde{\chi}_1^0 b q\bar{q}'$ and $\tilde{\chi}_1^0 b \bar{l} \nu_l$
($l=e,\mu,\tau$), as a function of $\Delta m$. 
Among the various Feynman diagrams
contributing to the decay, the dominant contribution arises from the
first diagram in Fig.~\ref{fig:4body}, with the virtual top-quark
and $W$ exchange. This is because the squark mixing matrix
elements, entering the $\tilde{u}_1-u_j-\tilde{\chi}_1^0$ coupling,
have larger values in the diagonal entries ({\it i.e.}~here for $j=3$),
and because of the smaller top-quark mass compared to the 
chargino and charged Higgs boson masses, which amount to several
hundred GeV in our scenarios.\footnote{In
  \cite{Boehm:1999tr} the most important contribution was due to the
  diagram with the virtual chargino and $W$ boson exchange, because
  smaller chargino masses were considered in the numerical analysis.}
The branching ratios for the final states involving $\bar{e} \nu_e$ and $\bar{\mu}
\nu_\mu$, marked by the green points, lie on top of each other. The
only difference in these final states arises from the diagrams with
virtual sleptons (last row in Fig.~\ref{fig:4body}), which are
negligibly small. 
In Fig.~\ref{fig:4bodydominant} (left) we see that for most parameter
points the four-body decay is not important and we again observe widespread 
results in the investigated parameter space as a consequence of the
smaller flavour symmetry. For the $U(3)$ symmetry
this is not the case, {\it cf.}~Fig.~\ref{fig:4bodydominant} (right), and a clear
hierarchy of the final states can be read off in the large $\Delta m$
region. The $\tilde{\chi}_1^0 b q\bar{q}'$ final state makes up $\sim 66$\% of the
four-body decay branching ratio, the $\tilde{\chi}_1^0 b \bar{l} \nu_l$
($l=e,\mu,\tau$) final states each contribute $\sim 11$\% which corresponds to
the branching ratios of an on-shell $W$ boson into quark and lepton
final states, respectively. In the threshold region due to the
non-vanishing $\tau$ mass, which is taken into account in our
calculation, the rise for the final state involving $\bar{\tau} \nu_\tau$
sets in later than for the decays with $\bar{e} \nu_e$ and $\bar{\mu} \nu_\mu$
final states. 

\subsection{The Stop Total Width and Branching Ratios and 
  Phenomenological Implications}
\begin{figure}[h!]
\begin{center}
\vspace*{-0.2cm}
\hspace*{-0.5cm}
\includegraphics[width=7.9cm]{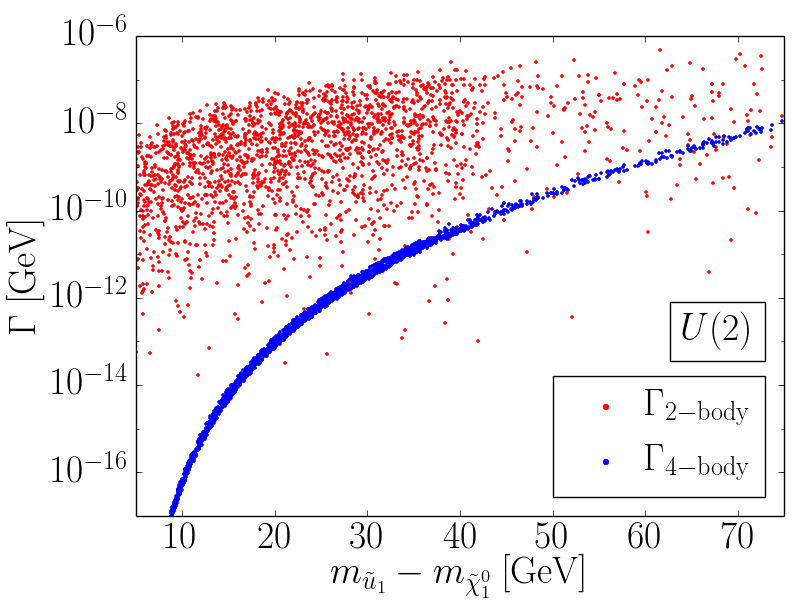} \hspace*{0.1cm}
\includegraphics[width=7.9cm]{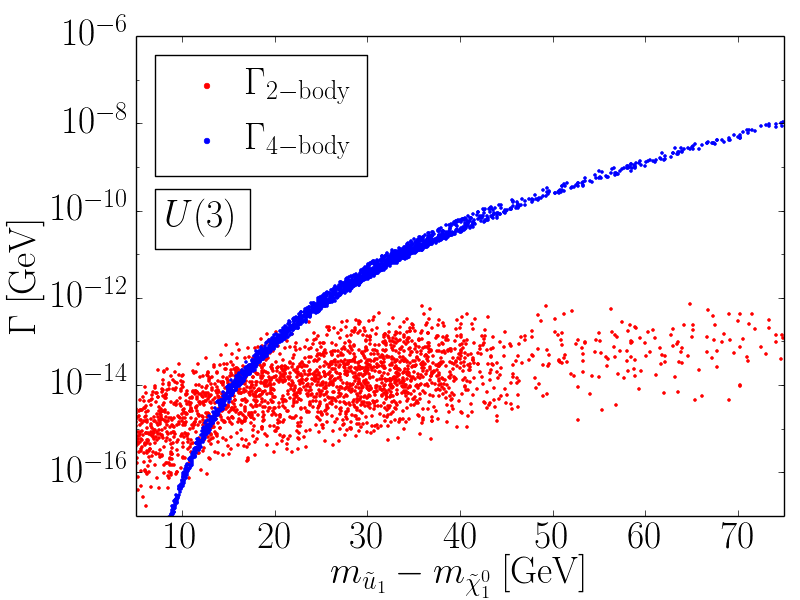} \\
\hspace*{-0.5cm}
\includegraphics[width=7.9cm]{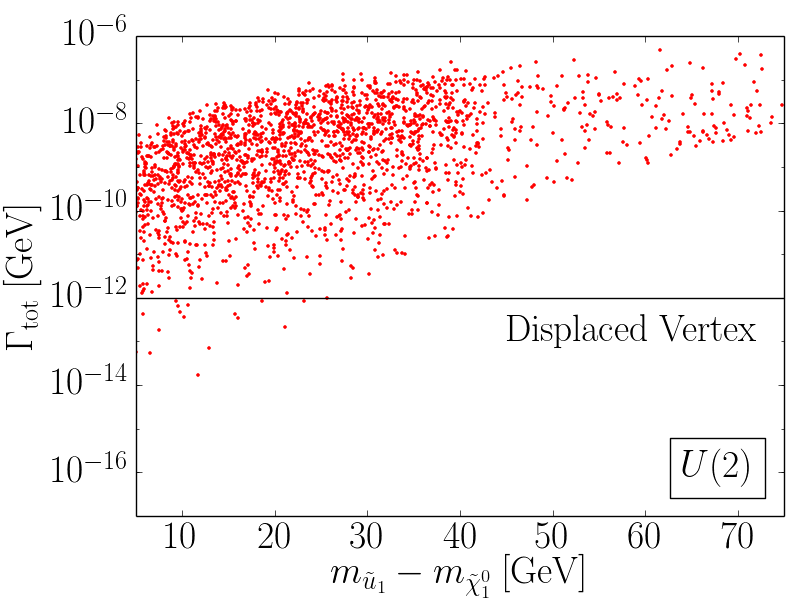} \hspace*{0.1cm}
\includegraphics[width=7.9cm]{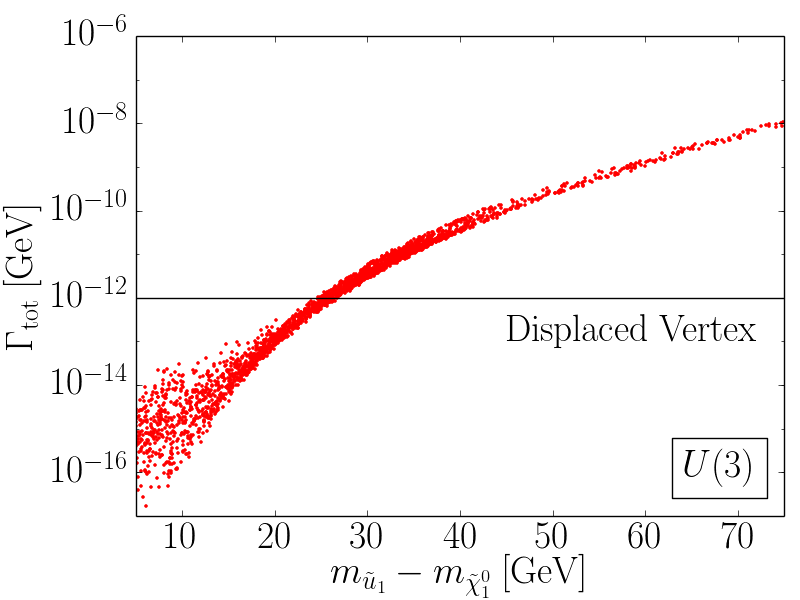} \\
\hspace*{-0.5cm}
\includegraphics[width=7.9cm]{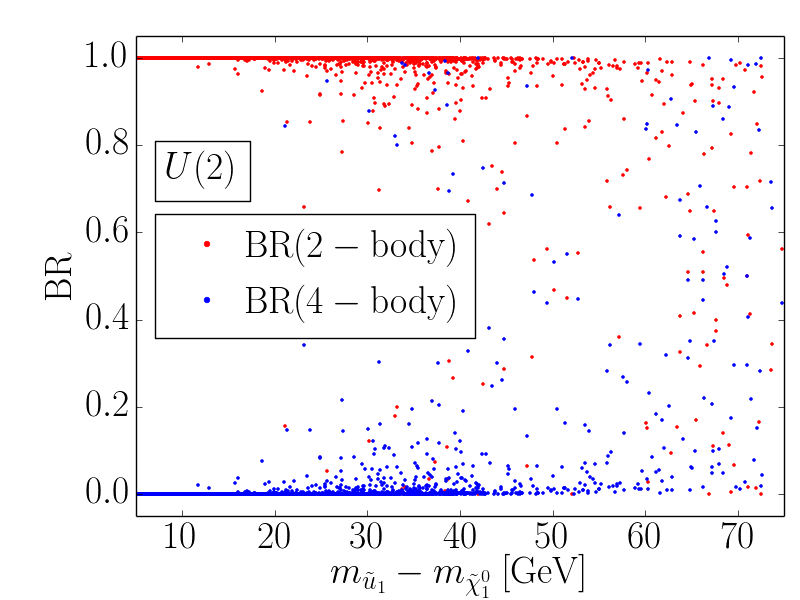} \hspace*{0.1cm}
\includegraphics[width=7.9cm]{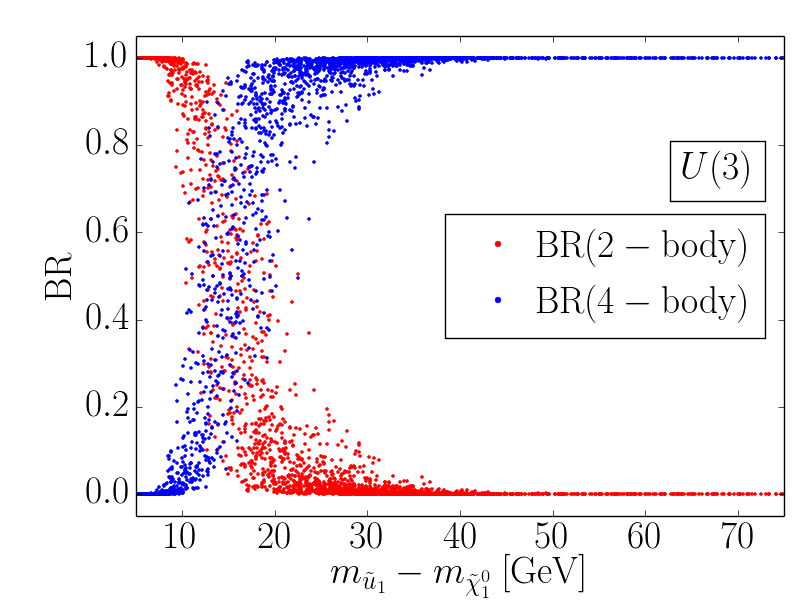} 
\caption{The two- and four-body decays widths (upper), the total
  widths (middle) and the branching ratios (lower) as a function of
  $\Delta m = m_{\tilde{u}_1}-m_{\tilde{\chi}_1^0}$, applying a $U(2)$
  (left) and a $U(3)$ (right) symmetry in the left-handed squark
  sector. \label{fig:totalwidths}}   
\end{center}
\end{figure}

\begin{figure}[h!]
\begin{center}
\vspace*{-0.2cm}
\hspace*{-0.5cm}
\includegraphics[width=8.1cm]{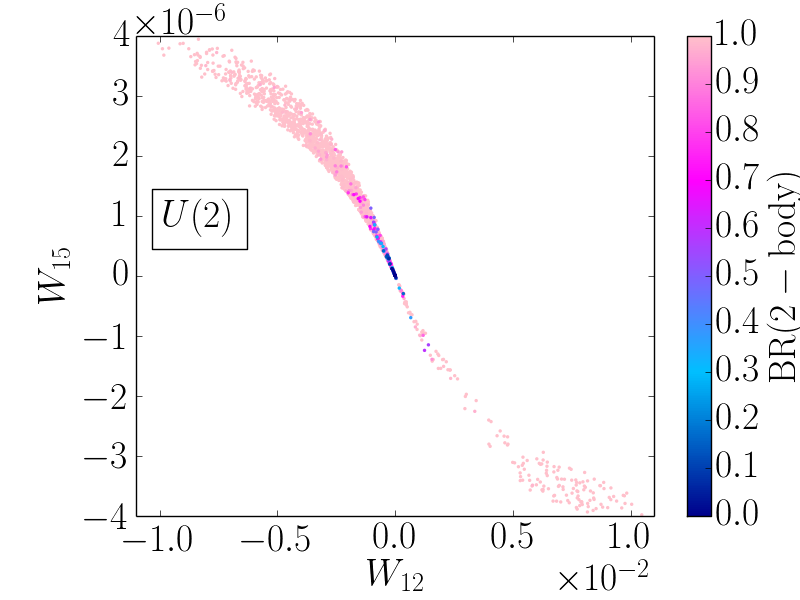} \hspace*{0.1cm}
\includegraphics[width=8.1cm]{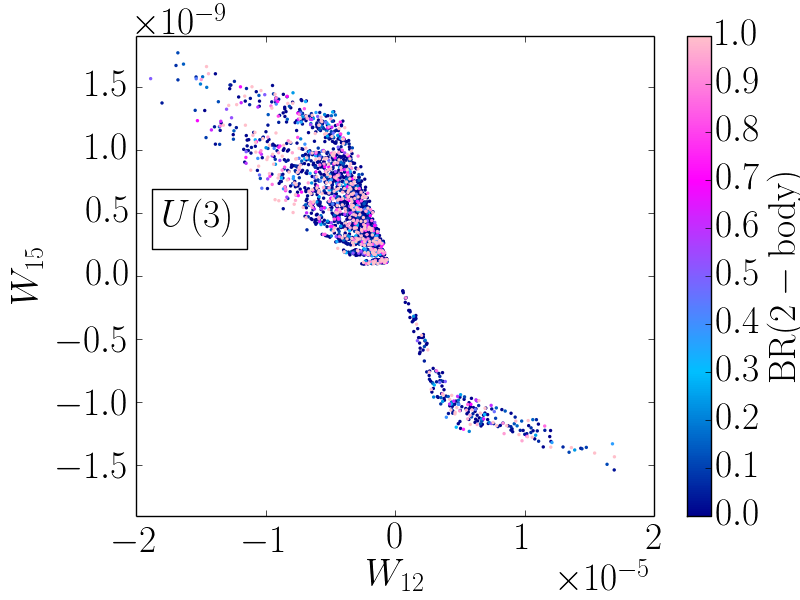} 
\caption{Values of the squark mixing matrix elements $W_{12}$ and
  $W_{15}$ for the points of the parameter scan passing all applied
  constraints. The colour code indicates the corresponding values of the branching
  ratios of the FCNC two-body decay assuming $U(2)$ (left) and $U(3)$
  (right) symmetry in the left-handed squark sector. 
\label{fig:matrixelements}}   
\end{center}
\end{figure}

In Fig.~\ref{fig:totalwidths} (upper) the partial two-body and
four-body decay widths are displayed for the two chosen flavour
symmetries. As can be inferred from the figures the results for the
two-body decay width scatter much more than for the four-body
decay, and even more in case of the smaller flavour symmetry $U(2)$,
Fig.~\ref{fig:totalwidths} (upper left). This is a consequence of the
former being mediated exclusively by FCNC couplings while the latter also
contains flavour conserving diagrams. In case of the $U(3)$ symmetry,
the off-diagonal squark mixing matrix elements $W_{12}$ and $W_{15}$
for the charm admixture to the top-flavour state, entering the
$\tilde{u}_1-c-\tilde{\chi}_1^0$ coupling, are much smaller, 
typically by three orders of magnitude, than if $U(2)$ is the applied
symmetry. This leads to a corresponding two-body FCNC decay width which
is about six orders of magnitudes smaller, {\it
  cf.}~Fig.~\ref{fig:totalwidths} (upper right). This is also
illustrated in Fig.~\ref{fig:matrixelements}, which shows the possible
values of $W_{12}$ and $W_{15}$ for the $U(2)$ and the $U(3)$
symmetry and the corresponding value of the FCNC decay
branching ratio, given by the colour code. As expected, in both cases
the right-chiral scharm admixture to the right-chiral stop-like squark
(given by $W_{15}$) is much smaller than the left-chiral scharm
admixture ($W_{12}$). Overall due to the larger flavour symmetry, for
the case of $U(3)$ shown in Fig.~\ref{fig:matrixelements} (right) the
mixing matrix elements $W_{12}$ and $W_{15}$ are ${\cal O} (10^3)$
smaller than for an assumed $U(2)$ symmetry, leading to a much smaller
two-body decay branching ratio compared to
Fig.~\ref{fig:matrixelements} (left), where we have branching ratios close to
one for the major part of the parameter points. \s 

The four-body decay width
is dominated by the diagrams mediated by flavour-conserving couplings,
so that it hardly depends on the details of the assumed flavour
symmetries and both for $U(2)$ and $U(3)$ yields values of ${\cal O}(10^{-8})$
GeV for $\Delta m =75$~GeV. Due to the smallness of the two-body
decay, it becomes the dominating decay channel already for mass
differences $\Delta m \gsim 20$~GeV for the enhanced flavour symmetry,
while for $U(2)$ the dominating decay for most parameter points is the
FCNC two-body decay over large parts of $\Delta m$. The decay widths
become comparable for $\Delta m \gsim 60$~GeV. The determination of the
relative size of the two decay channels to each other could hence 
be used to reveal information on the underlying flavour symmetry. \s

The total widths given by the sum of the two- and four-body decays in
the investigated $\Delta m$ range are depicted in
Fig.~\ref{fig:totalwidths} (middle). Dominated by the FCNC
decay, for the $U(2)$ symmetry it reaches almost $10^{-6}$~GeV for
$\Delta m =75$~GeV 
and the values are widely spread in the investigated mass
range. Applying the $U(3)$ symmetry, the values are spread for $\Delta
m \lsim 20$~GeV where the total width is dominated by the FCNC decay
and reaches maximum values of $10^{-8}$~GeV given by the four-body
decay width at $\Delta m =75$~GeV, see Fig.~\ref{fig:totalwidths}
(middle right). The black line at $\Gamma_{\text{tot}}= 10^{-12}$~GeV
corresponds to the value of the total width where displaced vertices can be
observed. It corresponds to a 
$\tilde{u}_1$ lifetime of the order of pico-seconds, which is a   
flight-time for the squark, that is long enough to lead to displaced vertices
in the detector.\footnote{For small decay widths the squark can
  hadronise before it decays. We did not take into account any long
  distance effects from hadronisation. Since we consider the inclusive
decay, the long distance effects can be estimated to be of ${\cal
  O}(\Lambda_{\text{QCD}}/m_{\tilde{u}_1})$ or even ${\cal
  O}(\Lambda^2_{\text{QCD}}/m_{\tilde{u}_1}^2)$, if the energy release
in the decay is much larger than $\Lambda_{\text{QCD}} \approx
200$~MeV, which is the scale where QCD becomes perturbative. See {\it
  e.g.}~Refs.~\cite{Donoghue:1995na,Hurth:2003vb,Paz:2010wu} with a
similar argument for rare $B$ decays.} Obviously, the more the FCNC couplings are
suppressed, the smaller is the total width, so that the observation of
displaced vertices allows for conclusions on the flavour symmetry of
the model as has been pointed out in
\cite{Hiller:2008wp,Hiller:2009ii}. In case the total decay width
is below $10^{-12}$~GeV even two displaced vertices could be possible,
one from the $\tilde{u}_1$ decay and the second from the $b$-quark
final state. \s

\begin{figure}[h]
\begin{center}
\hspace*{-0.5cm}
\includegraphics[width=10cm]{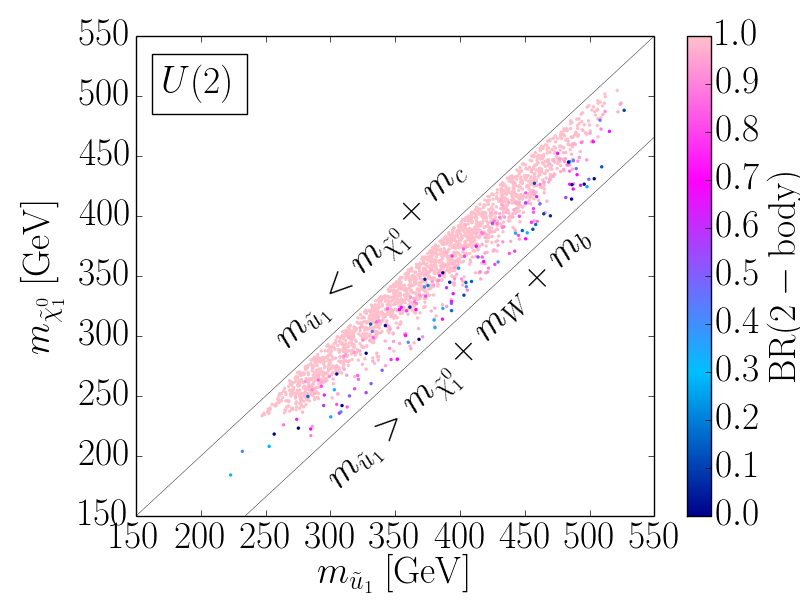} \\[0.5cm]
\hspace*{-0.5cm}
\includegraphics[width=10cm]{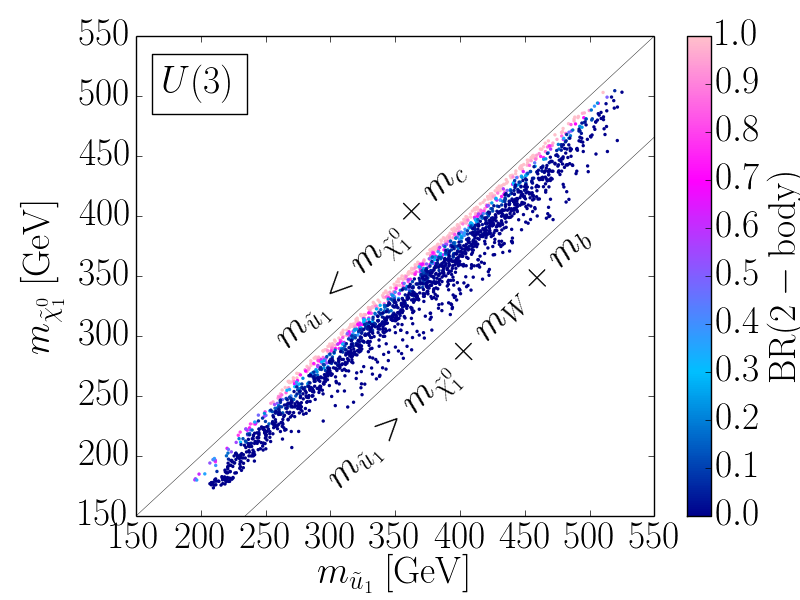} 
\caption{Parameter points of the scan, surviving all applied
  constraints, in the $m_{\tilde{\chi}_1^0}- m_{\tilde{u}_1}$
  plane. The colour code indicates the corresponding values of the
  FCNC two-body decay branching ratios. The upper grey line shows the
  threshold for the two-body decay, the lower grey line the threshold
  for the $\tilde{u}_1$ three body decay into $\tilde{\chi}_1^0 W
  b$. Upper: $U(2)$, lower: $U(3)$ flavour symmetry applied in the
  left-handed squark sector. \label{fig:inmassplane}}   
\end{center}
\end{figure}

The branching ratios finally, are displayed in
Fig.~\ref{fig:totalwidths} (lower). In case of the smaller flavour
symmetry, the branching ratio into $c \tilde{\chi}_1^0$ is close to
one for $\Delta m \lsim 20$~GeV. Beyond this value, however, the four-body
decay becomes important and both branching ratios can significantly
deviate from one, as can be inferred from Fig.~\ref{fig:totalwidths} (lower
left).\footnote{Taking this into account, the prospects for the FV two-body decay mode at the LHC have been investigated in \cite{Belanger:2013oka}. 
  The role of the four-body decay in the light stop mass
  window has been high-lighted in \cite{Delgado:2012eu}.} 
For the $U(3)$ symmetry there is a transition region $10 \mbox{
GeV} \lsim \Delta m \lsim 35 \mbox{ GeV}$, where the two-body and
four-body decay branching ratios cross, leading to a branching
close to one for the four-body decay above this $\Delta m$ range. In
this transition region, however, again the branching ratios for the two final
state signatures can deviate significantly from one. This
demonstrates that over large parts of the parameter space the
assumption of 100\% decay probability in either of the final states is
not valid. This therefore has to be taken into account by the
experiments by allowing also for deviations from one in the branching
ratios in the interpretation of their data. As evident from
Fig.~\ref{fig:stopexclusion} this has an important phenomenological
impact, as smaller branching ratios lead to considerably weakened
exclusion bounds on the lightest squark mass, {\it i.e.}~the lightest
stop mass. In order to further illustrate this, we show in
Fig.~\ref{fig:inmassplane} the values of the two-body decay branching
ratios for our investigated scenarios in the
$m_{\tilde{\chi}_1^0}-m_{\tilde{u}_1}$ plane in the region where the
two- and four-body decays are relevant. \s

Displayed are the points that result from our parameter scan and that 
survive the constraints, described in detail in
Section~\ref{sec:parameterscan}. In particular, the stop mass
exclusion limits from the LHC experiments have been applied 
in our refined approach, where deviations of the branching ratios from
one are taken into account, {\it cf.}~Fig.~\ref{fig:stopexclusion}. In
case of the $U(2)$ symmetry (upper plot) there are barely any viable parameter
points for $m_{\tilde{u}_1} \lsim 270$~GeV. In the scenarios, where
the two-body decay dominates this is due to the stop mass exclusion
bounds. In case the four-body decay is important (corresponding to the
blue points in the plot) it is either the mass bounds or the
constraints from the relic density, which exclude the points. As the
mass exclusions based on the four-body decay are weaker, there are
points that survive the constraints. This is also why in case
of the $U(3)$ symmetry (lower plot), where the four-body decay dominates in large
parts of the parameter space, there is a considerable amount of points
down to $\sim 205$~GeV. However, close to the two-body decay
threshold, {\it i.e.}~the upper grey line, the two-body decay becomes
more important and the more stringent mass exclusions based on this decay
apply, so that there are fewer allowed points. In this region of the
mass plane, our constraint on the relic density is fulfilled due to stop
co-annihilation. Close to the three-body decay threshold, however,
points are excluded due to the restrictions from the relic density. In
this range near the 
lower grey line above $\sim 300$~GeV neutralino annihilation via Higgs
boson exchange becomes effective, so that the constraint on the relic density
can be fulfilled and there are somewhat more points. This also applies in the
$U(2)$ case. The plots show in particular, that contrary to the naive 
application of the LHC exclusion limits, given by the full lines in
the plots of Fig.~\ref{fig:stopexclusion}, there are viable parameter points for
masses below these lines  both in the $U(2)$ and even more in the
$U(3)$ scenario. Thus, in the $U(2)$ scenario, where the two-body
decay dominates, there are points below 290~GeV (given by the limits from the
searches in the $c \tilde{\chi}_1^0$ final state) down to about
223~GeV. And in the $U(3)$ scenario, where $\tilde{u}_1$ mostly decays
into the four-body final state, masses below the limit given from the
four-body final state searches, {\it i.e.}~$\sim 270$~GeV, down to
approximately 195~GeV are allowed. This is because the assumption of a
two- or four-body decay branching ratio close to one, as applied by the
experiments, is not valid.\s 

Overall the picture for the branching ratios is as follows. 
For the smaller flavour symmetry $U(2)$ the 
dominating decay is the two-body FCNC decay with branching ratios
close to one, implying strongly suppressed branching ratios into the
four-body final state. However, for larger mass differences between
$m_{\tilde{u}_1}$ and $m_{\tilde{\chi}_1^0}$, close to the lower grey
line, the four-body decay becomes increasingly important and the
displayed branching ratio differs from one for a considerable amount
of parameter points, {\it   cf.}~in Fig.~\ref{fig:inmassplane} (upper) the
dark-pink to dark-blue points. For the enhanced flavour symmetry
$U(3)$ the situation evidently is 
reversed. In large parts of the $\Delta m$ region the four-body decay
dominates implying suppressed branching ratios for the FCNC
decay. For small $\Delta m$ values, {\it i.e.}~close to the upper grey
line, the FCNC decay, however, takes 
over, and branching ratios close to one are possible, as can be inferred
from Fig.~\ref{fig:inmassplane} (lower). From the plots for both
flavour symmetries it is evident, that the assumption of
branching ratios of one for either the two- or four-body decay is wrong
over large parts of the parameter space. Taking this into account and
re-interpreting the exclusion limits given by the
experiments accordingly, the exclusion bounds on the lightest squark, {\it
  i.e.}~the light stop, are significantly weakened. This should therefore be taken
into account in order to properly interpret the experimental data, in
particular at the next run of the LHC where a more extended part of
the low stop mass range will be probed. For
realistic, non-fine-tuned scenarios, there is hence still plenty of
room in the SUSY world for a light stop. 

\section{\label{sec:concl} Conclusions}
The supersymmetric partners of the top quark, the stops, play an
important role in the phenomenology of SUSY extensions. 
Taking into account constraints from Higgs
physics, $B$-physics and relic density measurements, light stops are
still allowed by the LHC experiments. The direct searches for
the lightest stop-like squark $\tilde{u}_1$ in the low mass range are
based on signatures from the two-body decay $\tilde{u}_1 \to c
\tilde{\chi}_1^0$ and from the four-body decay $\tilde{u}_1 \to
\tilde{\chi}_1^0 b f \bar{f}'$, which are the relevant decay channels
in the low stop mass region  $m_{\tilde{\chi}_1^0} + 
m_c \lsim m_{\tilde{u}_1} \lsim m_{\tilde{\chi}_1^0}+m_W + m_b$. 
We have revisited these two decay channels with the aim of providing 
precise theoretical predictions and subsequently investigating the implications
for the exclusion limits on the stop mass. \s

Allowing for FCNC couplings already at tree-level, we have calculated
for the first time the SUSY-QCD corrections to the two-body decay of the
lightest stop-like $\tilde{u}_1$ into charm and neutralino. They turn
out to be important, increasing the partial decay width by close to 27\% near
the kinematic threshold and approaching a constant value of about $5-7$\% far
above. In the calculation of the four-body decay we have taken into
account the contributions from the additional diagrams due to 
FCNC couplings and the finite masses of the final state
bottom quark and $\tau$ lepton. Both effects have not been available 
in the literature so far. Evidently, in the threshold region the mass
effects play an important role, but also far above they are still
significant, changing the partial width by up to 8\%. Above the bottom and
$\tau$ mass threshold the four-body decay is mainly given by the
flavour-conserving diagrams and hence much less sensitive to the
details of the flavour symmetry of the squark sector than the
two-body decay. We have assumed two different flavour patterns in the
left-handed squark sector, based on a $U(2)$ and a $U(3)$
symmetry. Accordingly, in the more symmetric $U(3)$ case the
flavour off-diagonal mixing in the squark sector is smaller, leading to
smaller FCNC two-body decay widths, while the four-body decay is
mostly insensitive to the flavour pattern. Depending on the flavour symmetry
the relative importance of the two- and four-body branching ratios to
each other changes, so that the knowledge on the branching ratios
gives information on the underlying flavour symmetry. In particular,
for the $U(3)$ case the total $\tilde{u}_1$ width can be very small leading to
displaced vertices in the detector, and even the observation of two
displaced vertices may be possible, from the $\tilde{u}_1$ decay and
from the $b$-quark final state. \s

The detailed investigation of the size of the two- and four-body decay 
branching ratios in our extensive parameter scan, which takes into account
all relevant constraints, reveals that the assumption of
branching ratios of one for either of the decay channels, which the 
experiments make in their exclusion plots for the lightest stop
quark, is wrong for large parts of the parameter space. Taking into
account information given by the experiments we have re-examined the
exclusions as function of the exact value of the $\tilde{u}_1$ branching ratio. As
expected, the bounds on the excluded lightest stop masses are
considerably weakened. Applying this information on the scenarios of
our parameter scan we find that there is still a sizeable amount of
scenarios with allowed stop mass values below the presently given experimental
exclusion limits. This is in particular the case for scenarios where the
four-body decay dominates, {\it i.e.}~for the $U(3)$ symmetry
assumption, as here the exclusions given by the experiment are
weaker. \s

In summary, the precise prediction for the FCNC two-body and the
four-body decay of the lightest stop-like squark in the low stop mass
region, taking into account SUSY-QCD corrections, mass effects and
flavour violation at tree-level, as done here for the first time, is
indispensable for the correct interpretation of the experimental
exclusion limits. Deviations of either of the branching ratios from
one in large parts of the parameter space considerably weaken the stop
exclusion limits. This should be taken into account at the next run of
the LHC with higher center-of-mass energy and luminosity where a bigger 
part of the stop mass region dominated by these two decay channels
will be probed. Contrary to the present naive picture, in the SUSY world
there is actually a larger range of light stop masses that is still
allowed by the LHC experiments.  
 
\section*{Appendix}
\section*{ A The NLO Decay Width \boldmath$\tilde{u}_1\to 
c\tilde{\chi}_1^0$ \label{app:stop2body}}
In this Appendix, we give the result for the SUSY-QCD corrected NLO
decay width of the FCNC decay. We give the finite part of the
result that remains after the application of the renormalisation
procedure and performing the sum of the virtual and real
corrections. Furthermore, we assume $m_c$ to be zero. 
The decay width at NLO is composed of, {\it cf.}~also
Eq.~(\ref{eq:nlowidth}), 
\beq
\Gamma^{\text{NLO}} = \Gamma^{\text{LO}} + \Gamma^{\text{virt}} +
\Gamma^{\text{real}} + \Gamma^{\text{CT}} \;, \label{appstop1}
\eeq
with $\Gamma^{\text{LO}}$ as given in Eq.~\eqref{eq:lowidth}. 
The scalar integrals appearing in the virtual and counterterm
contributions, $\Gamma^{\text{virt}}$ and $\Gamma^{\text{CT}}$, are defined as 
\begin{align}
A(m_1^2) &= \f{(2\pi\mu)^{4-D}}{i\pi^2}\int \mathrm{d}^D
q\f{1}{
\left(q^2-m_1^2\right)}\label{appstop2}\\
 B_{i}(m_1^2, m_2^2) &=\f{(2\pi\mu)^{4-D}}{i\pi^2}\int \mathrm{d}^D
q\f{1}{
\left(q^2-m_1^2\right)(\left(q+p_i\right)^2-m_2^2)}\label{appstop3}\\
p_{\mu} B_1(p^2,m_1^2, m_2^2)&=\f{(2\pi\mu)^{4-D}}{i\pi^2}\int \mathrm{d}^D
q\f{q_{\mu}}{
\left(q^2-m_1^2\right)(\left(q+p\right)^2-m_2^2)}\\
 C_{ij}(m_1^2, m_2^2, m_3^2) &=\f{(2\pi\mu)^{4-D}}{i\pi^2}\int \mathrm{d}^D
q\f{1}{
\left(q^2-m_1^2\right)(\left(q+p_i\right)^2-m_2^2)(\left(q+p_i+p_j\right)^2-m_3^
2)}\,.
\end{align}
The indices $i,j$ of the four-momenta refer to either $c$, $\tilde{u}_1$ or 
$\tilde{\chi}_1^0$ and will be specified later. 
Applying on-shell renormalisation the virtual corrections
$\Gamma^{\text{virt}}$ only receive contributions from the gluon and gluino
vertex corrections, $\Gamma^{\text{virt}}_g$  and
$\Gamma^{\text{virt}}_{\tilde{g}}$, depicted in Fig.~\ref{fig:virtualcorr}
(upper),
\beq
\Gamma^{\text{virt}}= \Gamma^{\text{virt}}_g +
\Gamma^{\text{virt}}_{\tilde{g}} \;,
\eeq 
with the specific contributions given by
\begin{align}
 \Gamma^{\text{virt}}_ g=&\frac{-\alpha_s(\mu)}{24\pi^2 m_{\tilde{u}_1}^3} 
\left((g^L _{211})^2+(g^R _{211})^2\right)
\text{Re}\left[ B_{\tilde{\chi}_1^0}(0 , m_{\tilde{u}_1}^2)(2 m_{\tilde{\chi}_1^0}^4 - 
2 m_{\tilde{\chi}_1^0}^2 m_{\tilde{u}_1}^2) \right.\nonumber\\ 
- & 2 B_c(0,0) (m_{\tilde{\chi}_1^0}^2 - m_{\tilde{u}_1}^2)^2 -
  B_{\tilde{u}}(0,m_{\tilde{u}_1}^2)
 (m_{\tilde{\chi}_1^0}^4 - m_{\tilde{u}_1}^4)\nonumber \\
+&\left. 2 C_{c, \tilde{\chi}_1^0}(0,0, m_{\tilde{u}_1}^2)
(m_{\tilde{\chi}_1^0}^2 -m_{\tilde{u}_1}^2)^3\right] 
\end{align}
and
\begin{align}
\Gamma^{\text{virt}}_{\tilde{g}}=&\frac{\alpha_s(\mu)}{12 
\pi^2\,m_{\tilde{u}_1}}\left(1-\frac{m_{\tilde{\chi}_1^0}^2}{m_{\tilde{u}_1}^2}
\right)\sum_{i=1}^3\sum_{s=1}^6 \text{Re} \left[m_{\tilde{\chi}_1^0}
B_{\tilde { \chi }_0 }( m_{u_i}^2, m_{\tilde{u}_s}^2) \left[ \right.\right.\nonumber
\\ 
- &  W^*_{1i} W^*_{s5}\left(g^L_{211}
  g^L_{is1}m_{u_i} +g^L_{211}g^R_{is1}m_{\tilde{\chi}_1^0} \right) - W^*_{1\,i+3}W^*_{s2} 
\left(g^R_{211} g^R_{is1}m_{u_i} +
  g^R_{211}g^L_{is1}m_{\tilde{\chi}_1^0} \right) \nonumber \\+& \left.
 W^*_{1i} W^*_{s2} \, g^R_{211}g^R_{is1}m_{\tilde{g}}+ W^*_{1\,i+3}
 W^*_{s5} \, g^L_{211}g^L_{is1}m_{\tilde{g}}\right]
   \nonumber \\ 
+ &B_{\tilde{u}}(m_{u_i}^2, m_{\tilde{g}}^2)\left[W^*_{1i}W^*_{s5}
 \left(g^L_{211}g^L_{is1}m_{\tilde{\chi}_1^0} m_{u_i} +
g^L_{211}g^R_{is1}m_{\tilde{u}_1}^2\right)\right. \nonumber \\
+&  W^*_{1\,i+3} W^*_{s2} \left(g^R_{211}g^R_{is1}m_{\tilde{\chi}_1^0} 
m_{u_i}+g^R_{211}g^L_{is1}m_{\tilde{u}_1}^2 \right) \nonumber \\
-& \left.W^*_{1\,i+3}
  W^*_{s5}\, g^L_{211}g^L_{is1}m_{\tilde{\chi}_1^0} m_{\tilde{g}} 
 - W^*_{1i} W^*_{s2} \,g^R_{211}g^R_{is1}m_{\tilde{\chi}_1^0}  m_{\tilde{g}}\right]\nonumber\\
+ & C_{c,\tilde{\chi}_1^0}(m_{\tilde{g}}^2,m_{\tilde{u}_s}^2,m_{u_i}^2)\left[-
W^*_{1i}W^*_{s5} \left( g^L_{211}g^L_{is1}m_{u_i}m_{\tilde{\chi}_1^0}
 (m_{\tilde{g}}^2 -
 m_{\tilde{u}_s}^2)\right.\right.\nonumber\\
+&\left.g^L_{211}g^R_{is1}(m_{\tilde { \chi } _0 } ^2 m_{\tilde{g}}^2
  - m_{\tilde{u}_1}^2 m_{\tilde{u}_s}^2)  \right)-
 W^*_{1\,i+3}W^*_{s2} \left( g^R_{211}g^R_{is1}m_{u_i}m_{\tilde{\chi}_1^0}
 (m_{\tilde{g}}^2 - m_{\tilde{u}_s}^2)\right.\nonumber\\
+&\left.g^R_{211}g^L_{is1}(m_{\tilde { \chi } _0 } ^2 m_{\tilde{g}}^2
  -   m_{\tilde{u}_1}^2 m_{\tilde{u}_s}^2) \right)+W^*_{s5}
W^*_{1\,i+3}\left(g^L_{211}g^L_{is1}m_{\tilde{\chi}_1^0}  m_{\tilde{g}} 
(m_{\tilde{\chi}_1^0}^2+ m_{\tilde{g}}^2 \right.\nonumber
\\ - &\left. m_{\tilde{u}_1}^2 - m_{\tilde{u}_s}^2)+g^L_{211}g^R_{is1}m_{\tilde{g}}m_{u_i} 
(m_{\tilde{\chi}_1^0}^2-m_{\tilde{u}_1} ^2)\right)+
W^*_{s2} W^*_{1i} \left(g^R_{211}g^R_{is1}m_{\tilde{\chi}_1^0}
 m_{\tilde{g}} \right.\nonumber\\ \quad
&\left.\left.\left. (m_{\tilde{\chi}_1^0}^2+ m_{\tilde{g}}^2- m_{\tilde{u}_1}^2 - m_{\tilde{u}_s}^2)
+g^R_{211}g^L_{is1}m_{\tilde{g}}m_{u_i}
(m_{\tilde{\chi}_1^0}^2-m_{\tilde{u}_1}^2)\right) \right]
\right] \; . 
\end{align}
The gluino mass is denoted by $m_{\tilde{g}}$ and $\mu$ denotes the 
renormalisation scale. The decay width stemming from the counterterms,
$\Gamma^{\text{CT}}$, reads
\begin{equation}
 \Gamma^{\text{CT}}=\frac{1}{8\pi} 
m_{\tilde{u}_1}\left(1-\frac{m^2_{\tilde{\chi}_1^0}}{m_{\tilde{u}_1}^2}
\right)^2\left(\delta g^L_{211} g^L_{211}+\delta g^R_{211} g^R_{211}\right)\,,
\end{equation}
with $\delta g^{L/R}$ as defined in Eqs.~\eqref{eq:deltagl} and
\eqref{eq:deltagr}. For the computation of the counterterms the
quark and squark self-energies are needed. The squark self-energy
can be cast into the form ($s,t=1,...,6$) 
\begin{align} \nonumber
\tilde{\Sigma}_{st}(p^2)=&\frac{\alpha_s(\mu)}{3\pi}\sum_{r=1}^6\left[A(m_{
\tilde { u }_r } ^2) \sum_{i=1}^3 \sum_{j=1}^3
(W^*_{si} W_{ri} - W^*_{s \, i+3} W_{r \, i+3})(W^*_{rj}
W_{tj} - W^*_{r \, j+3} W_{t \, j+3}) \right]
\\-&\frac{2\alpha_s(\mu)}{3\pi}\sum_{i=1}^3\left[A(m_{u_i}^2)(W^*_{s\,
    i+3} W_{t\,i+3}+W^*_{si} W_{ti})+ A(m_{\tilde{g}}^2)(W^*_{s\, i+3}
  W_{t\,i+3} + W^*_{si} W_{ti}) \right.
\nonumber\\
\quad &\quad\quad\quad\quad\;\;\; +B_{p^2}(m_{u_i}^2, 
m_{\tilde{g}}^2)\left((m_{u_i}^2+m_{\tilde{g}}^2-p^2)(W^*_{si}W_{ti}+
W^*_{s\,i+3}W_ { t\,i+3})\right.
\nonumber\\
\quad&\left.\left.\quad\quad\quad\quad\;\; -2 
m_{u_i} m_{\tilde{g}} 
(W^*_{si}W_{t\,i+3}+W^*_{s\,i+3}W_ { ti})\right)
\right ]
\nonumber\\
+&\frac{\alpha_s(\mu)}{3\pi}\delta_{st}\left[A(m_{\tilde{u}_s}^2)-2(p^2+m_{
    \tilde{u}_s}^2) B_{p^2}(0,m_{\tilde{u}_s}^2)\right ]\,.  
\end{align}
The symbol $B_{p^2}$ is defined in analogy to Eq.~\eqref{appstop3}, with 
unspecified momentum $p^2$. The quark self-energies read
\begin{align}
 \Sigma^L_{ij}&=-\frac{2\alpha_s(\mu)}{3\pi}\sum_{s=1}^6 
B_1(p^2,m_{\tilde{g}}^2, 
m_{\tilde{u}_s}^2)W_{si}W^*_{sj}\\
 \Sigma^R_{ij}&=-\frac{2\alpha_s(\mu)}{3\pi}\sum_{s=1}^6 
B_1(p^2,m_{\tilde{g}}^2, 
m_{\tilde{u}_s}^2)W_{s\,i+3}W^*_{s\,j+3}\\
\Sigma^{Ls}_{ij}&=-\frac{2\alpha_s(\mu)}{3\pi}\sum_{s=1}^6 m_{\tilde{g}} 
B_{p^2}(m_{\tilde{g}}^2, 
m_{\tilde{u}_s}^2)W^*_{s\,i+3}W_{s\,j}\\
\Sigma^{Rs}_{ij}&=-\frac{2\alpha_s(\mu)}{3\pi}\sum_{s=1}^6 m_{\tilde{g}} 
B_{p^2}(m_{\tilde{g}}^2, 
m_{\tilde{u}_s}^2)W_{s\,j+3}W^*_{s\,i}\,.
\end{align}
For the computation of the real corrections we use the parametrisation
as in \cite{Campbell:2004ch},
\beq
r^2 & \equiv & \frac{(p_{\tilde{u}_1}-p_c-p_g)^2}{m_{\tilde{u}_1}^2} =
\frac{m_{\tilde{\chi}_1^0}^2}{m_{\tilde{u}_1}^2} \\
p_c p_g &=& \frac{m_{\tilde{u}_1}^2}{2} (1-r)^2 y \\
p_{\tilde{u}_1} p_g &=& \frac{m_{\tilde{u}_1}^2}{2} (1-r^2) (1-z) \;,
\eeq
in terms of the four-momenta of the squark, charm quark and gluon,
$p_{\tilde{u}_1}$, $p_c$ and $p_g$, respectively. The squared matrix
element evaluated from the Feynman diagrams of the real corrections,
depicted in Fig.~\ref{fig:real}, is integrated over the three-particle
phase space in $D=4-2\epsilon$ dimensions, with $\epsilon \equiv
\epsilon_{\text{IR}}$. The $D$-dimensional differential
three-particle phase space $d\Phi^{(3)}$ reads
\beq
d\Phi^{(3)} (p_c,p_{\tilde{\chi}_1^0},p_g;p_{\tilde{u}_1}) &=& d
\Phi^{(2)} (p_c,p_{\tilde{\chi}_1^0}; p_{\tilde{u}_1})
\frac{(1-r)^2}{16 \pi^2} \frac{(m_{\tilde{u}_1}^2)^{1-\epsilon} (4
  \pi)^\epsilon}{\Gamma (1-\epsilon)} \left( \frac{1+r}{1-r}\right)^{2
\epsilon} \nonumber \\
&& \int_0^1 dz (r^2 + (1-r^2) z )^{-\epsilon} \int_0^{y_{max}} dy \,
y^{-\epsilon} (y_{max} - y)^{-\epsilon} \;,
\eeq
where $d\Phi^{(2)}$ denotes the differential two-particle phase space,
$p_{\tilde{\chi}_1^0}$ the four-momentum of the neutralino and
$\Gamma$ the Gamma function. The upper integration limit $y_{max}$ is given by
\beq
y_{max} = \frac{(1+r)^2 z (1-z)}{(z-r^2 z+r^2)} \;.
\eeq  
This leads to the following result for the finite part of the real corrections,
\begin{align}
\Gamma^{\text{real}}&=-\frac{\alpha_s(\mu)}{288\pi^2} 
m_{\tilde{u}_1}((g^L_{211})^2 + (g^R_{211})^2)
\left[10\pi^2-99+204 r^2-20\pi^2 r^2
\right.\nonumber\\\nonumber&-105 r^4 + 10 \pi^2 r^4 - 6(-1 + r^2)^2 
\log^2\left(\f{m_{\tilde{u}_1}^2}{\mu^2}\right)+ 
24 r^2 \log(r^2) - 18 r^4 \log(r^2) \\
&+ 
 60 \log(1 - r^2) - 120 r^2 \log(1 - r^2) + 
60 r^4 \log(1 - r^2) - 24 \log(1 - r^2)^2 \nonumber\\
&+ 
48 r^2 \log^2(1 - r^2) - 
24 r^4 \log^2(1 - r^2) - 
 6(-1 + r^2)^2 \log\left(\f{m_{\tilde{u}_1}^2}{\mu^2}\right)
 \nonumber\\
&\left. \quad \left(-5 + 4 \log(1 - r^2)\right) - 
24 (-1 + r^2)^2 Li_2(1-r^2)
 \right]\,,
\end{align}
with the Spence function 
\begin{equation}
Li_2(z)=-\int_0^z du\frac{\log(1-u)}{u} \hspace*{1.5cm} , \; z\in \mathbb C 
\setminus[1,\infty)\,.
\end{equation}

\vspace*{0.5cm}
\section*{Acknowledgments}
We greatly acknowledge helpful discussions with M. Hohlfeld. 
RG appreciates discussions with G.~Arcadi, L.~Di Luzio and
T.~Hermann. 
RG and EP thank Mathieu Pellen for discussions on the real
corrections. 
AW thanks T.~Chwalek, M.~Mozer and Z.~Rurikova for discussions. 
RG acknowledges partial financial support from the
``Landesgraduiertenf\"orderung (LGK)'' and from the Graduiertenkolleg ``GRK
1694: Elementarteilchenphysik bei h\"ochster Energie und h\"ochster
Pr\"azision''.
RG (partly) and AW have been supported by the ``Karlsruhe School of
Elementary Particle and Astroparticle Physics: Science and Technology
(KSETA)''.
The work by RG and MM was supported in part by the Deutsche
Forschungsgemeinschaft via the ``Sonderforschungsbe\-reich/Transregio
SFB/TR-9 Computational Particle Physics''. 

\vspace*{0.5cm}

\end{document}